# Dynamic surface critical behavior of systems with conserved bulk order parameter: Detailed RG analysis of the semi-infinite extensions of model $B$ with and without nonconservative surface terms*

F. Wichmann and H. W. Diehl

*Fachbereich Physik, Universität - Gesamthochschule - Essen, D-45117 Essen*

## Abstract

The dynamic surface critical behavior of macroscopic systems whose dynamic bulk critical behavior is described by model $B$ is investigated. The semi-infinite extensions of bulk model $B$ introduced in a previous treatment [Phys. Rev. B **49**, 2846 (1994)] and called models $B_A$ and $B_B$ are studied in detail by means of field-theoretic RG methods. The distinctive feature of these models is the presence or absence of relevant nonconservative surface terms. The earlier results on the structure of the required counterterms, the flow equations, and the representation of the surface critical exponents of dynamic quantities at the ordinary and special phase transitions in terms of static bulk and surface exponents are corroborated by means of an explicit two-loop calculation for $4-\epsilon$ dimensions. For parameter values corresponding to a given static surface universality class, models $B_A$ and $B_B$ represent distinct dynamic surface universality classes. Differences between these manifest themselves in

*Dedicated to Prof. Dr. Herbert Wagner, Ludwig-Maximilians-Universität München, on the occasion of his 60th birthday.

the behavior of scaling functions for dynamic surface susceptibilities. RG-improved perturbation theory is used to compute some of these susceptibilities to one-loop order.

## I. INTRODUCTION

The past two decades have witnessed enormous activity and progress in the field of critical phenomena at boundaries, such as surfaces, interfaces, or walls [1–4]. This has led to the development of a fairly satisfactory theory for static phenomena of this kind in thermal equilibrium systems. As far as quantitative predictions are concerned, the theory has not quite yet reached the degree of accuracy of the theory of bulk critical phenomena. However, practically all the sophisticated quantitative techniques of the modern theory of critical phenomena have been extended to systems with boundaries. Thus there are no fundamental reasons that would prevent one from achieving the same numerical precision as in the bulk case. The quality of quantitative results that can be attained through theoretical analyses is mainly determined by how much effort one is willing to spend [5–8].

An equally important and impressive development has taken place on the experimental side. During the past decade powerful surface-sensitive techniques have been developed and refined [3]. As a consequence, accurate experimental studies of static critical phenomena at surfaces and their quantitative comparison with theoretical predictions have become possible.

The situation is much less satisfactory if one considers *dynamic* critical phenomena at surfaces. Despite the high level of sophistication of the available experimental techniques, the quantitative study of such dynamic surface critical behavior remains a major challenge. On the theoretical side, the pertinent knowledge accumulated is rather limited: only a few semi-infinite extensions of known bulk models [18] have been examined in some detail [9–17], and reliable quantitative results are scarce. Furthermore, from a more fundamental point of view one would like the theory to achieve a classification of the possible types of dynamic critical behavior at surfaces into corresponding *dynamic surface universality classes* and their representation in terms of appropriate prototype continuum models. This analog of the program expounded in Ref. [18] for dynamic bulk critical behavior has only just been started.

In a previous paper [17], hereafter referred to as I, the dynamic surface critical behavior of semi-infinite systems with the following properties was studied:

(i) Their dynamic bulk critical behavior is represented by model $B$ of Ref. [18].

(ii) Their static surface critical behavior is described by the semi-infinite $\phi^4$ model of Ref. [2].

Because of (i), the order parameter $\phi$ must be a conserved density away from the surface, but it need not have this property in the vicinity of the surface. Such a local violation of the continuity equation was found to correspond to a *relevant* surface perturbation whose strength could be parametrized by a surface variable $\tilde{c}_0$. The corresponding extensions of model $B$ to the $d$-dimensional half-space $z \geq 0$ for the cases $\tilde{c}_0 > 0$ with, and $\tilde{c}_0 = 0$ without, nonconservative surface terms were called models $B_A$ and $B_B$, respectively. Both represent distinct dynamic surface universality classes. Hence each one of the static universality classes in question — namely, those of the ordinary, special, and extraordinary transitions [2] — splits up into two distinct dynamic ones.

In I, models $B_A$ and $B_B$ were investigated using field-theoretic renormalization group (RG) methods in $4-\epsilon$ dimensions. Invoking a combination of arguments, the general structure of the necessary counterterms was derived. This led to the assertion that the required renormalization functions could all be expressed in terms of known ones. In turn, this implied that the dynamic (bulk and surface) critical indices can all be written in terms of known static (bulk and surface) exponents.

The aim of the present paper is twofold: to verify the findings of I by explicitly carrying through the field-theoretic RG program to two-loop order, and to compute quantities that discriminate between the dynamic surface universality classes represented by models $B_A$ and $B_B$. Obvious candidates for such quantities are the universal scaling functions associated with dynamic surface susceptibilities. Using RG-improved perturbation theory, we have calculated such scaling functions to one-loop order for the case of the special transition. Owing to the complicated form of the free dynamic propagators, these calculations are rather demanding. This has kept us from computing more scaling functions to the same order of RG-improved perturbation theory.

The remainder of the paper is organized a follows. In Sec. II we briefly recapitulate the definition of models $B_A$ and $B_B$ and their functional-integral representation. In Sec. III we describe details of the calculations needed to verify the renormalization of the theory to two-loop order. In Sec. IV our one-loop results for susceptibility scaling functions are presented. This includes a discussion of their crossover from critical to hydrodynamic behavior. Section V contains a brief discussion of our results and conclusions. In Appendices A–F various calculational details are explicated.

## II. DEFINITION AND FUNCTIONAL-INTEGRAL REPRESENTATION OF MODELS $B_A$ AND $B_B$

We consider a time-dependent $n$-component order-parameter field $\phi(\mathbf{x},t) \equiv (\phi_\alpha(\mathbf{x},t))$ on the half-space $V = \{\mathbf{x} = (\mathbf{x}_\parallel, z) \mid \mathbf{x}_\parallel \in \mathbb{R}^{d-1},\ z \geq 0\}$ bounded by the surface $\partial V = \{(\mathbf{x}_\parallel, z = 0) \mid \mathbf{x}_\parallel \in \mathbb{R}^{d-1}\}$. As expounded in I, the models we are concerned with are defined by the Langevin equation

$$\dot{\phi}_\alpha(\mathbf{x},t) \;=\; -\lambda_0 \left( D \frac{\delta \mathcal{H}}{\delta \phi_\alpha} \right)(\mathbf{x},t) + \zeta_\alpha(\mathbf{x},t)\ . \tag{1}$$

Here $\lambda_0$ is a constant, and

$$\mathcal{H} = \int_V d^dx \left[ \tfrac{1}{2} (\nabla\phi)^2 + \tfrac{1}{2}\, \tau_0\, \phi^2 + \tfrac{1}{4!}\, u_0\, \phi^4 + \delta(z)\, \tfrac{1}{2} c_0\, \phi^2 \right] \tag{2}$$

is the Hamiltonian of the semi-infinite $\phi^4$ model. The operator $D$ is defined through its action

$$(Df)(\mathbf{x}) = -\Delta f(\mathbf{x}) \tag{3}$$

on functions $f$ over $V$ satisfying the boundary condition

$$\partial_n f = \tilde{c}_0\, f \tag{4}$$

on $\partial V$, where the derivative $\partial_n$ is along the inner normal. This Robin boundary condition [19] guarantees that the self-adjoint extension of $D$ exists; it is satisfied by the functions

$$\mathcal{H}_\phi(\mathbf{x},t) \equiv \left[\left(-\Delta + \tau_0 + \tfrac{1}{6}\, u_0\, \phi^2\right)\phi\right](\mathbf{x},t) \tag{5}$$

produced by the functional derivative $\delta\mathcal{H}/\delta\phi$ in (1).

The Gaussian noise $\zeta$ has mean zero and variance

$$\langle \zeta_\alpha(\mathbf{x},t)\, \zeta_\beta(\mathbf{x}',t')\rangle = 2\lambda_0\, \delta_{\alpha\beta}\, D_{\mathbf{x}\mathbf{x}'}\, \delta(t-t')\ , \tag{6}$$

where $D_{\mathbf{x}\mathbf{x}'}$ is the integral kernel associated with $D$. Writing it in the formally symmetric form

$$D_{\mathbf{x}\mathbf{x}'} = \overleftarrow{\nabla}\delta(\mathbf{x}-\mathbf{x}')\overrightarrow{\nabla}' + \tilde{c}_0\, \delta(z)\, \delta(\mathbf{x}-\mathbf{x}')\ , \tag{7}$$

one sees that the variable $\tilde{c}_0$ measures indeed the strength of the nonconservative surface terms.

The above equations with $\tilde{c}_0 > 0$ and $\tilde{c}_0 = 0$ define models $B_A$ and $B_B$, respectively.

In order to apply renormalized perturbation theory, it is convenient to transform to the equivalent functional-integral representation. According to I, this gives the dynamic action [17]

$$\begin{aligned}\mathcal{J} = \int dt\Bigg( &\int_V \left[\tilde{\phi}\,\dot{\phi} - \lambda_0(\Delta\tilde{\phi})(\mathcal{H}_\phi - \tilde{\phi})\right] \\ &+ \lambda_0 \int_{\partial V} \left\{\left[(\partial_n - \tilde{c}_0)\tilde{\phi}\right]\left[\mathcal{H}_\phi - \tilde{\phi} - \delta(0)(\partial_n - c_0)\phi\right] + (\Delta\tilde{\phi})(\partial_n - c_0)\phi\right\}\Bigg)\ ,\end{aligned} \tag{8}$$

in which $\tilde{\phi}$ is the usual auxiliary field [20]. The singularity $\propto \delta(0)$ is a boundary term that arises from the surface potential $\propto \delta(z)$ in (2) upon integration by parts. To avoid it, one can replace the latter $\delta$-function by an appropriate regularized one, such as

$$\delta_B(z) \equiv B\, e^{-Bz} \tag{9}$$

with arbitrarily large but finite $B$. Then $\delta(0)$ gets replaced by $B$. Since the boundary conditions found in I,

$$\partial_n \phi = c_0 \phi \ , \quad \partial_n \mathcal{H}_\phi = \tilde{c}_0 \mathcal{H}_\phi \tag{10a}$$

$$\partial_n \Delta\tilde{\phi} = c_0 \Delta\tilde{\phi} \ , \quad \partial_n \tilde{\phi} = \tilde{c}_0 \tilde{\phi} \ , \tag{10b}$$

imply that the contributions from the surface integral in (8) vanish, we do not have to worry any further about the singular term $\propto \delta(0)$ in the action $\mathcal{J}$.

## III. RENORMALIZATION

As in I, we consider the general response and correlation functions

$$\begin{aligned} &W^{(\tilde{N},N;\tilde{M},M)}(\mathbf{x},\mathbf{r},t) \\ &\equiv \left\langle \prod_{j=1}^{\tilde{N}} \Big( -\lambda_0 \Delta\tilde{\phi}\Big) \prod_{k=1}^{N} \phi \prod_{l=1}^{\tilde{M}} \Big( -\lambda_0 \Delta\tilde{\phi}|_s\Big) \prod_{m=1}^{M} \phi|_s \right\rangle^C \ , \end{aligned} \tag{11}$$

as well as the related functions

$$\begin{aligned} &G^{(\tilde{N},N;\tilde{M},M)}(\mathbf{x},\mathbf{r},t) \\ &\equiv \left\langle \prod_{j=1}^{\tilde{N}} \tilde{\phi} \prod_{k=1}^{N} \phi \prod_{l=1}^{\tilde{M}} \tilde{\phi}|_s \prod_{m=1}^{M} \phi|_s \right\rangle^C \ . \end{aligned} \tag{12}$$

Here the superscript $C$ indicates cumulant averages. The variables $\mathbf{x}$, $\mathbf{r}$, $t$ represent the $N+\tilde{N}$ points off the surface, the $M+\tilde{M}$ parallel coordinates of the surface points, and the set of all time arguments, respectively.

According to the arguments given in I, to renormalize these functions in $4-\epsilon$ dimensions, it should be sufficient to supplement the usual static reparametrizations [2]

$$\phi = Z_\phi^{1/2} \phi^R \ , \quad \tau_0 = \mu^2 Z_\tau \tau + \tau_b \ , \tag{13a}$$

$$u_0 = \frac{\mu^\epsilon}{s_d} Z_u u \ , \quad s_d = (4\pi)^{-d/2} \ , \tag{13b}$$

$$\phi|_s = (Z_\phi Z_1)^{1/2} (\phi|_s)^R \ , \quad c_0 = \mu \, Z_c \, c + c_{\rm sp} \ , \tag{13c}$$

and the well-known dynamic ones

$$\tilde{\phi} = Z_\phi^{-1/2} \tilde{\phi}^R \ , \quad \lambda_0 = \mu^{-4} \, Z_\phi \, \lambda ; \, , \tag{14a}$$

by

$$\tilde{\phi}|_s = Z_\phi^{-1/2} (\tilde{\phi}|_s)^R \ , \quad \tilde{c}_0 = \mu \tilde{c} \ , \tag{15a}$$

$$(\Delta \tilde{\phi})_s = (Z_1 / Z_\phi)^{1/2} \, [(\Delta \tilde{\phi})_s]^R \ . \tag{15b}$$

Thus the renormalized functions [21]

$$G_R^{(\tilde{N},N;\tilde{M},M)} = Z_\phi^{\frac{\tilde{N}+\tilde{M}-N-M}{2}} Z_1^{-\frac{M}{2}} \, G^{(\tilde{N},N;\tilde{M},M)} \tag{16}$$

and

$$W_R^{(\tilde{N},N;\tilde{M},M)} = Z_\phi^{-\frac{\tilde{N}+\tilde{M}+N+M}{2}} Z_1^{-\frac{\tilde{M}+M}{2}} \, W^{(\tilde{N},N;\tilde{M},M)} \tag{17}$$

should be ultraviolet-finite. Utilizing dimensional regularization, we will verify this to the order of two-loops for the functions $G_R^{(0,1;1,0)}$ and $W_R^{(0,1;1,0)}$. To this end we shall first compute these functions for $\tau = c = 0$. In a second step we shall then check the cancellation of $c$-dependent poles by expanding about the special point. For notational simplicity we will set the number of components $n = 1$ henceforth.

### A. Graphical notation

The Feynman graphs of the functions we are interested in involve the free propagators

$$G(\mathbf{x}, t; \mathbf{x}', t') = \langle \phi(\mathbf{x}, t) \, \tilde{\phi}(\mathbf{x}', t') \rangle_f \tag{18}$$

and

$$C(\mathbf{x}, t; \mathbf{x}', t') = \langle \phi(\mathbf{x}, t) \, \phi(\mathbf{x}', t') \rangle_f^C \ . \tag{19}$$

It is favorable to work in a $\mathbf{p}z\omega$ representation. For their Fourier transforms, defined by

$$\hat{G}(\mathbf{p}; z, z'; \omega) \equiv \int d^{d-1}x_{\|} \int dt\, e^{-i[\mathbf{p}\cdot(\mathbf{x}_{\|}-\mathbf{x}'_{\|})-\omega(t-t')]}\, G(\mathbf{x}, t; \mathbf{x}', t')\ , \tag{20}$$

we use the graphical notation

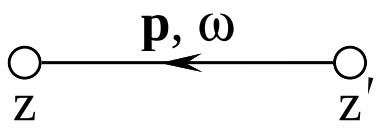

$$= \hat{G}(\mathbf{p}; z, z'; \omega) \tag{21}$$

and

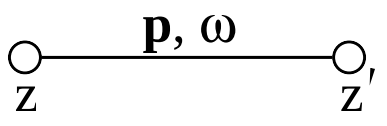

$$= \hat{C}(\mathbf{p}; z, z'; \omega)\ . \tag{22}$$

In order to indicate whether external points, such as the above ones labelled $z$ and $z'$, are off or on the surface, we use open circles (as shown) for points with $z > 0$ and crossed circles (as in Fig. 1) for surface points.

The explicit expression for the response propagator (21) may be gleaned from I; since it is rather lengthy, even in the special case $\tau = c = 0$, we have relegated it to Appendix A [see Eq. (A1)].

The correlation propagator $\hat{C}$ can be expressed in terms of $\hat{G}$, exploiting the fluctuation-dissipation theorem. One has

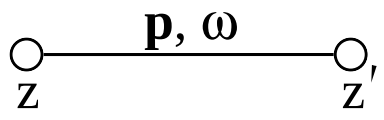

$$= \frac{2\lambda_0}{\omega}\, \mathrm{Im}\Big[\ \Big]$$

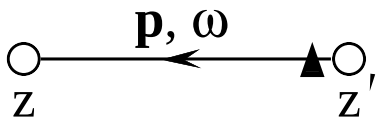

$$= \frac{2}{\omega}\, \mathrm{Im}\big[\chi^{[0]}(\mathbf{p}; z, z'; \omega)\big]\ , \tag{23}$$

where the full triangle stands for the negative Lapacian $\mathbf{p}^2 - \partial_z^2$, and $\chi^{[0]}$ is a zero-loop susceptibility.

In conformity with these conventions, we represent the four-point vertex of the action (8) by

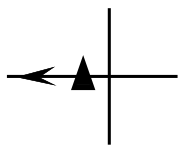

$$. \tag{24}$$

As a further graphical element, we shall need the vertex representing insertions of the operator $-\lambda_0 \int dt \int_{\partial V} \phi \Delta \tilde{\phi}$ to which $c_0$ couples. We use the symbol

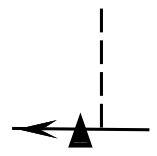 . (25)

## B. Calculation for $c = 0$

In Fig. 1 the graphs of $G^{(0,1;1,0)}$ are depicted up to two-loop order. The corresponding graphs of $W^{(0,1;1,0)}/\lambda_0$ are obtained from these by putting a full triangle on one external leg.

To compute the latter graphs, it is convenient to proceed as follows. We amputate the free propagator lines attached to the external points. The resulting amputated graphs must then be considered as distributions with respect to their dependence on $z$ and $z'$. Regularized by analytic continuation in $\epsilon$, these distributions are well-defined and can be Laurent expanded in $\epsilon$ [22]. For simple powers such as $z^{-2+\epsilon}$, the Laurent expansion is well-known and can be looked up in Ref. [22]. However, in general one must study the action of the distributions on test functions to determine their $\epsilon$ expansion. To illustrate the procedure, let us consider the integral

$$\int_0^\infty dz \, \mathcal{D}_\epsilon(z) \, f(z) \; , \tag{26}$$

where $\mathcal{D}_\epsilon(z)$ is a distribution and $f(z)$ a test function. As a test function, $f(z)$ can be written in terms of its Laplace transform $\tilde{f}(\alpha)$ as

$$f(z) = \int\limits_{C-i\infty}^{C+i\infty} d\alpha \, e^{-\alpha z} \tilde{f}(\alpha) \; , \tag{27}$$

with arbitrary $C > 0$. Hence it is sufficient to know how $\mathcal{D}_\epsilon$ acts on the class of functions $e^{-\alpha z}$ with $\mathrm{Re}\,\alpha > 0$. Upon computing the corresponding (dimensionally regularized) integrals, one arrives at Laurent series of the form

$$\int_0^\infty dz \, \mathcal{D}_\epsilon(z) \, e^{-\alpha z} = \sum_{j=-n_0}^{\infty} \frac{d^{(j)}(\alpha)}{\epsilon^j} \; , \tag{28}$$

where $n_0$ is the order of the highest pole. In the next step one has to find distributions $\mathcal{D}^{(j)}(z)$ that solve

$$\int_0^\infty dz\, \mathcal{D}^{(j)}(z)\, e^{-\alpha z} = d^{(j)}(\alpha)\ . \tag{29}$$

This yields the desired expansion

$$\mathcal{D}_\epsilon(z) = \sum_{j=-n_0}^{\infty} \frac{\mathcal{D}^{(j)}(z)}{\epsilon^j}\ . \tag{30}$$

In Appendix A the $\epsilon$ expansion of the amputated two-point graphs resulting from the graphs of Fig. 1 are given to the required order in $\epsilon$. From these results the corresponding $\epsilon$ expansion of each $W$ and $G$-function with $N + \tilde{N} + M + \tilde{M} = 2$ can be obtained in a straightforward fashion by applying the distributions to the respective products of free propagators that the amputated external legs represent. The explicit expressions one obtains for the graphs of $G^{(0,1;1,0)}$ and $W^{(0,1;1,0)}$ are given in Appendix B.

To determine the renormalized functions, we substitute these results into (16) and (17) and use the reparametrizations (13a)–(15b) together with the known two-loop results [2]

$$Z_\phi = 1 - \frac{1}{12\epsilon}u^2 + O(u^3)\ , \tag{31}$$

$$Z_u = 1 + \frac{3}{\epsilon}u + O(u^2)\ , \tag{32}$$

$$Z_1 = 1 + \frac{1}{\epsilon}u + \left[\frac{2}{\epsilon^2} - \frac{1}{\epsilon}\right]u^2 + O(u^3) \tag{33}$$

for $n = 1$. Proceeding in this fashion, we have explicitly verified to two-loop order the cancellation of all dimensional poles in $G_R^{(0,1;1,0)}$ and $W_R^{(0,1;1,0)}$. Details of these lengthy calculations are given in Appendix C. Their final results read

$$G_R^{(0,1;1,0)}(\mathbf{p}, z, \omega; \lambda, \tilde{c}, u) = \frac{1}{\lambda(k_+^2 - k_-^2)(2k_+k_- + \tilde{c}(k_+ + k_-))}\Big\{2k_+\, e^{-k_- z} - 2k_-\, e^{-k_+ z}$$
$$- \tfrac{1}{4}u\Big[\mathcal{E}_+\, e^{-k_+ z} - \mathcal{E}_-\, e^{-k_- z}\Big] + u^2\, O(\epsilon^0) + O(u^3)\Big\} \tag{34}$$

and

$$W_R^{(0,1;1,0)}(\mathbf{p}, z, \omega; \lambda, \tilde{c}, u) = \frac{1}{2k_+k_- + \tilde{c}(k_+ + k_-)}\Big\{\Big[(k_- + \tilde{c})e^{-k_+ z} + (k_+ + \tilde{c})e^{-k_- z}\Big]$$

$$+\tfrac{1}{4}u\Big(\Big[C_E(k_-+\tilde{c})-\tfrac{1}{2}\mathcal{F}_+\Big]e^{-k_+z}+\Big[C_E(k_++\tilde{c})-\tfrac{1}{2}\mathcal{F}_-\Big]e^{-k_-z}\Big)$$

$$+u^2\,O(\epsilon^0)+O(u^3)\Big\}\;. \tag{35}$$

Here $C_E$ is Euler's number, and the quantities $\mathcal{E}_\pm$, $\mathcal{F}_\pm$, and $k_\pm$ are defined by Eqs. (B3) and (C5), by (B10) and (C11), and by (C3) of Appendices B and C, respectively.

### C. Expansion about the special point

Until now our calculations were restricted to the case of critical surface enhancement, $c = 0$. We now wish to check whether the chosen reparametrization also absorb the additional poles that appear for $c \neq 0$. In order to bypass the complicated computation of $c$-dependent graphs, we expand about the multicritical theory. Since the critical bare surface enhancement $c_{\rm sp}$ vanishes in our perturbative approach based on dimensional regularization, this amounts to an expansion in powers of $c_0$ for the bare theory. For simplicity, we will restrict ourselves to the case of model $B_B$ in this calculation, setting $\tilde{c} = 0$.

From the action (8) we see that the operation $-d/dc_0$ produces an insertion of the (Fourier transformed) operator $-\lambda_0[(\phi\Delta\tilde{\phi})_s]_{\mathbf{P},\Omega}$ at zero parallel momentum $\mathbf{P}$ and zero frequency $\Omega$. Let us denote the analogs of the functions $G^{(\tilde{N},N;\tilde{M},M)}$ with $I$ insertions of $-\lambda_0[(\phi\Delta\tilde{\phi})_s]_{\mathbf{P},\Omega}$ by $G^{(\tilde{N},N;\tilde{M},M;I)}$. Then the expansion of the functions $\hat{G}$ (or $\hat{W}$) in powers of $c_0$ can be written as

$$\hat{G}^{(\tilde{N},N;\tilde{M},M)}(c_0) = \hat{G}^{(\tilde{N},N;\tilde{M},M)}$$

$$-\,c_0\,\hat{G}^{(\tilde{N},N;\tilde{M},M;1)} + O(c_0^2)\;, \tag{36}$$

where the functions on the right-hand side are taken at $c_0 = 0$, and it is understood that the inserted operator has $\mathbf{P} = \mathbf{0}$ and $\Omega = 0$.

A word of caution must be added here: We wish to use this expansion in the massless case $\tau = 0$, where it becomes problematic because of infrared singularities [23]. Strictly speaking, we should take the limit $\mathbf{P} \to \mathbf{0}$ of the massless functions $\hat{G}^{(\tilde{N},N;\tilde{M},M;I)}$ only after the infrared singularities have been resummed via the RG [24]. For the graphs of $\hat{G}^{(\tilde{N},N;\tilde{M},M;1)}$ we are going to consider, setting $\mathbf{P}=\mathbf{0}$ causes no problems as their ultraviolet poles can be safely identified.

According to the reparametrizations (13a)–(15b), the renormalized $G$-functions with insertions should be given by

$$G_R^{(\tilde{N},N;\tilde{M},M;I)} = Z_\phi^{\frac{\tilde{N}+\tilde{M}-N-M}{2}} Z_1^{-\frac{M}{2}} Z_c^I \, G^{(\tilde{N},N;\tilde{M},M;I)} \ . \tag{37}$$

We will verify to the order of two loops that all ultraviolet poles cancel in $G_R^{(0,1;1,0;1)}$, if we use the previously given two-loop expressions for the renormalization factors together with the known $n=1$ result [24,2]

$$Z_c = 1 + \frac{1}{\epsilon}\, u + \left[\frac{2}{\epsilon^2} + \frac{1}{12\epsilon}(1-4\pi^2)\right] u^2 + O(u^3) \ . \tag{38}$$

The graphs of $G^{(0,1;1,0;1)}$ contributing to this order are shown in Fig. 2. Appendix D contains the $\epsilon$ expansions of the required distributions and of the graphs themselves. Upon making the transition to renormalized quantities and summing the contributions, one arrives at

$$\hat{G}_R^{(0,1;1,0;1)}(\mathbf{p},z,\omega;\lambda,\tilde{c}=0,u) = \frac{k_-\, e^{-k_+ z} + k_+\, e^{-k_- z}}{2\lambda\,(k_+ + k_-)\, k_+^2\, k_-^2} + u\, O(\epsilon^0) + u^2 O(\epsilon^0) + O(u^3) \ . \tag{39}$$

## IV. SCALING FUNCTIONS

According to (15a) $\tilde{c}_0$ is not renormalized. This implies the trivial flow behavior

$$\bar{\tilde{c}}(l) = \frac{\tilde{c}}{l} \tag{40}$$

under RG transformations $\mu \to \mu l$. On physical grounds the initial value $\tilde{c}$ must be non-negative. Hence there are two fixed-point values of physical interest: $\tilde{c}_A^* = \infty$ and $\tilde{c}_B^* = 0$. In the enlarged parameter space of the dynamic theory, each static fixed point (describing the ordinary, special, or extraordinary transition) unfolds into a line $\tilde{c} \geq 0$ between pairs of dynamic fixed points with these two values of $\tilde{c}$. The fixed points corresponding to models $B_A$ and $B_B$ are infrared-stable and infrared-unstable in the $\tilde{c}$-direction, respectively (see the flow diagram depicted in Fig. 5 of I).

The members of each of these pairs of fixed points represent, *distinct dynamic* surface universality classes with the *same static* critical behavior. Since the dynamic critical exponents can be expressed in terms of static ones, we must look at other universal properties to see the difference between these universality classes — and hence between the dynamic surface critical behavior of model $B_A$ and of model $B_B$.

As a simple illustrative quantity we will examine the surface susceptibility

$$\begin{aligned}\chi_{11}(\mathbf{r},t) &\equiv \left.\frac{\delta\langle\phi|_s(\mathbf{r},t)\rangle}{\delta h_{1,0}(\mathbf{0},0)}\right|_{h_{1,0}=0} \\ &= W^{(0,1;0,1)}(\mathbf{r},t;\mathbf{0},0)\end{aligned} \tag{41}$$

for $c = 0$ and $\tau \geq 0$. Here $h_{1,0}(\mathbf{r},t)$ is a position and time-dependent surface magnetic field.

From the general results for the $W$-functions given in Eqs. (3.67) and (3.69a)–(3.70) of I we can infer the scaling form of $\chi_{11}(\mathbf{r},t)$ for $\tau > 0$ and $c = 0$. (As usual, these scaling forms are obtained in our field-theoretic approach as fixed-point solutions to the RG equations for the *renormalized* functions, but they describe the critical behavior of the *bare* functions as well. We therefore drop the subscript $R$ in the rest of this section.)

Upon Fourier transformation we arrive at the scaling form

$$\hat{\chi}_{11}(\mathbf{p},\omega,\tau,\tilde{c}) \approx C_1\,\tau^{-\gamma_{11}^{\mathrm{sp}}}\,\hat{\Psi}(\mathsf{p},\mathsf{w},\tilde{\mathsf{c}}) \tag{42}$$

with

$$\mathsf{p} = C_2\,p\,\tau^{-\nu}\ , \quad \mathsf{w} = C_3\,\omega\,\tau^{-\nu\zeta}\ , \quad \tilde{\mathsf{c}} = C_4\,\tilde{c}\,\tau^{-\nu}\ . \tag{43}$$

Here $\gamma_{11}^{\mathrm{sp}} = \nu(1-\eta_{\|}^{\mathrm{sp}})$ is the usual surface susceptibility exponent of the special transition, $\zeta = 4 - \eta$ denotes the dynamic bulk exponent, and the $C_i$ are nonuniversal constants.

As is borne out by our perturbative results (see Appendix E and below), the left-hand side of (42) exists for $\tilde{c} = c_A^*$ and $\tilde{c} = c_B^*$. Hence the corresponding limiting values

$$\hat{\Psi}_{A,B}(\mathsf{p},\mathsf{w}) \equiv \hat{\Psi}(\mathsf{p},\mathsf{w},\tilde{\mathsf{c}}^*_{A,B}) \tag{44}$$

of the scaling function $\hat{\Psi}$ for $\tilde{\mathsf{c}} \to c^*_A \equiv \infty$ and $\tilde{\mathsf{c}} \to c^*_A \equiv 0$ should exist. Utilizing RG-improved perturbation theory, we have computed the scaling functions $\hat{\Psi}_A$ and $\hat{\Psi}_B$ to one-loop order. Some details of the calculation can be found in Appendix E. The results are

$$\begin{aligned}\hat{\Psi}_A(\mathsf{p},\mathsf{w}) = {} & \frac{2(\mathsf{k}_+ + \mathsf{k}_-)}{(\mathsf{k}_+ + \mathsf{k}_-)^2 + 1} + \frac{\epsilon}{12\mathsf{k}_+\mathsf{k}_-(\mathsf{k}_+ + \mathsf{k}_-)(\mathsf{k}_+ - \mathsf{k}_-)^2\left[(\mathsf{k}_+ + \mathsf{k}_-)^2 + 1\right]^2} \\ & \cdot\Big\{\mathcal{A} + \mathcal{B}_+ + \mathcal{B}_- + 4\mathsf{k}_+\mathsf{k}_-(\mathsf{k}_+^2 - \mathsf{k}_-^2)^2 + (1 - C_E)(\mathsf{k}_+ - \mathsf{k}_-)^2 \\ & \cdot\Big[(\mathsf{k}_+ + \mathsf{k}_-)^2\,(\mathsf{k}_+ + 10\mathsf{k}_+\mathsf{k}_- + \mathsf{k}_-^2) - 2(\mathsf{k}_+ + \mathsf{k}_-)^2 + 1\Big]\Big\} + O(\epsilon^2)\end{aligned} \tag{45}$$

and

$$\begin{aligned}\hat{\Psi}_B(\mathsf{p},\mathsf{w}) = {} & \frac{(\mathsf{k}_+ + \mathsf{k}_-)^2 - 1}{2\mathsf{k}_+\mathsf{k}_-(\mathsf{k}_+ + \mathsf{k}_-)} + \frac{\epsilon}{48\mathsf{k}_+^3\,\mathsf{k}_-^3\,(\mathsf{k}_+ - \mathsf{k}_-)^2(\mathsf{k}_+ + \mathsf{k}_-)^3} \\ & \cdot\Big\{\mathsf{k}_+\mathsf{k}_-\,\mathcal{A} + \mathsf{k}_+^2\,\mathcal{B}_+ + \mathsf{k}_-^2\,\mathcal{B}_- - 4\mathsf{k}_+^2\,\mathsf{k}_-^2\,(\mathsf{k}_+^2 - \mathsf{k}_-^2)^2 \\ & + (1 - C_E)(\mathsf{k}_+ - \mathsf{k}_-)^2\Big[(\mathsf{k}_+ + \mathsf{k}_-)^4\,(\mathsf{k}_+^2 - \mathsf{k}_+\mathsf{k}_- + \mathsf{k}_-^2) \\ & - 2(\mathsf{k}_+ + \mathsf{k}_-)^2\,(\mathsf{k}_+^2 + \mathsf{k}_+\mathsf{k}_- + \mathsf{k}_-^2) + (\mathsf{k}_+^2 + 3\mathsf{k}_+\mathsf{k}_- + \mathsf{k}_-^2)\Big]\Big\} + O(\epsilon^2)\ ,\end{aligned} \tag{46}$$

with

$$\begin{aligned}\mathcal{A} = 4\mathsf{k}_+\mathsf{k}_-\Big\{ & \sqrt{\frac{\mathsf{k}_+ + \mathsf{k}_- - 2}{\mathsf{k}_+ + \mathsf{k}_- + 2}}\,\ln\left[\frac{\sqrt{\mathsf{k}_+ + \mathsf{k}_- + 2} + \sqrt{\mathsf{k}_+ + \mathsf{k}_- - 2}}{\sqrt{\mathsf{k}_+ + \mathsf{k}_- + 2} - \sqrt{\mathsf{k}_+ + \mathsf{k}_- - 2}}\right]\Big[(\mathsf{k}_+ + \mathsf{k}_-)^2 \\ & + 2(\mathsf{k}_+ + \mathsf{k}_-) - (\mathsf{k}_+ - \mathsf{k}_-)^2\,(\mathsf{k}_+ + \mathsf{k}_-)^4 + 2\Big(2\mathsf{k}_+^3\mathsf{k}_-^2 + 2\mathsf{k}_+^2\mathsf{k}_-^3 - \mathsf{k}_+^5 - \mathsf{k}_+^4\mathsf{k}_- - \mathsf{k}_+\mathsf{k}_-^4 \\ & - \mathsf{k}_-^5\Big)\Big] + (2 - C_E)\Big[\mathsf{k}_-^6 + 2\mathsf{k}_-^5\mathsf{k}_+ - \mathsf{k}_-^4\mathsf{k}_+^2 - 4\mathsf{k}_-^3\mathsf{k}_+^3 - \mathsf{k}_-^2\mathsf{k}_+^4 + 2\mathsf{k}_-\mathsf{k}_+^5 + \mathsf{k}_+^6\Big]\Big\}\end{aligned} \tag{47}$$

and

$$\mathcal{B}_\pm = -4\mathsf{k}_\pm\mathsf{k}_\mp(\mathsf{k}_\mp^2 - 1)^{1/2}\,(\mathsf{k}_\pm + \mathsf{k}_\mp)\,(\mathsf{k}_\pm^2 - \mathsf{k}_\mp^2 - 1)^2\,\ln\left(\frac{\sqrt{\mathsf{k}_\mp + 1} + \sqrt{\mathsf{k}_\mp - 1}}{\sqrt{\mathsf{k}_\mp + 1} - \sqrt{\mathsf{k}_\mp - 1}}\right)\ , \tag{48}$$

where

$$\mathsf{k}_\pm = \sqrt{\mathsf{p}^2 + \frac{1}{2} \pm \sqrt{\frac{1}{4} + i\mathsf{w}}}\ . \tag{49}$$

A well-known general property of our model is that dynamic susceptibilities $\hat{\chi}(\omega, .)$ in the limit $\omega \to 0$ go over into their static counterparts $\hat{\chi}^{\rm stat}(.)$. Hence

$$\lim_{\omega\to 0} \hat{\chi}_{11}(\mathbf{p}, \omega, \tau, c, \tilde{c}) = \hat{\chi}_{11}^{\rm stat}(\mathbf{p}, \tau, c) \ . \tag{50}$$

This implies that as $\mathsf{w} \to 0$ with all other arguments fixed, the scaling function $\hat{\Psi}(\mathsf{p}, \mathsf{w}, \tilde{\mathsf{c}})$ — and hence both $\hat{\Psi}_A$ as well $\hat{\Psi}_B$ — must approach a limiting function that is independent of $\tilde{\mathsf{c}}$ and equal to the corresponding scaling function of $\hat{\chi}_{11}^{\rm stat}$. Setting again $c = 0$, we write

$$\hat{\chi}_{11}^{\rm stat}(\mathbf{p}, \tau) \approx C_1 \, \tau^{-\gamma_{11}^{\rm sp}} \, \Psi^{\rm stat}(\mathsf{p}) \ . \tag{51}$$

Then

$$\begin{aligned} \hat{\Psi}(\mathsf{p}, \mathsf{w}\!=\!0, \tilde{\mathsf{c}}) &= \hat{\Psi}_A(\mathsf{p}, \mathsf{w}\!=\!0) = \hat{\Psi}_B(\mathsf{p}, \mathsf{w}\!=\!0) \\ &= \Psi^{\rm stat}(\mathsf{p}) \ . \end{aligned} \tag{52}$$

As an explicit check the reader may verify that for $\mathsf{w} = 0$ the $\epsilon$ expansions (45) and (46) of $\hat{\Psi}_A$ and $\hat{\Psi}_B$ both reduce to the same result

$$\begin{aligned} \Psi^{\rm stat}(\mathsf{p}) = \frac{1}{\sqrt{\mathsf{p}^2+1}} + \frac{\epsilon}{3}\Bigg[ & \frac{2 - C_E}{2\sqrt{\mathsf{p}^2+1}} + \frac{1 - C_E}{4\,(\mathsf{p}^2+1)^{3/2}} \\ & - \frac{\mathsf{p}}{\mathsf{p}^2+1} \ln\left(\mathsf{p} + \sqrt{\mathsf{p}^2+1}\right) \Bigg] + O\left(\epsilon^2\right) \end{aligned} \tag{53}$$

of the static theory.

On the other hand, the functions $\hat{\Psi}_A$ and $\hat{\Psi}_B$ differ in general whenever $\mathsf{w} \neq 0$. Inspection of their $\mathsf{w}$-dependence reveals, in particular, that

$$\mathrm{Re}\left[\hat{\Psi}_A(\mathsf{p}\!=\!0, \mathsf{w}\!\to\!0)\right] - \hat{\Psi}^{\rm stat}(0) \sim \mathsf{w}^{3/2}, \tag{54}$$

but

$$\mathrm{Re}\left[\hat{\Psi}_B(\mathsf{p}\!=\!0, \mathsf{w}\!\to\!0)\right] - \hat{\Psi}^{\rm stat}(0) \sim \mathsf{w}^{1/2} \ . \tag{55}$$

To elucidate further these differences between the corresponding dynamic surface universality classes of models $B_A$ and $B_B$, we consider the behavior of $\chi_{11}$ as a function of $t$. Fourier transformation of (42) yields

$$\chi_{11}(\mathbf{p},t,\tau,\tilde{c}) \approx C_1 C_3^{-1}\,\tau^{\nu\zeta-\gamma_{11}^{\rm sp}}\,\Psi(\mathsf{p},\mathsf{t},\tilde{\mathsf{c}})$$
$$= C_1 C_3^{-1}\,\mathsf{t}^{-(\zeta-1+\eta_\parallel^{\rm sp})/\zeta}\,\Xi(\mathsf{p},\mathsf{t},\tilde{\mathsf{c}}) \tag{56}$$

with $\mathsf{t} \equiv C_3^{-1} t \tau^{\nu\zeta}$, where $\Psi(.,\mathsf{t},.)$ is the inverse Fourier transform of $\hat{\Psi}(.,\mathsf{w},.)$. Depending on whether $\tilde{\mathsf{c}}$ is small or large, we may set $\tilde{\mathsf{c}}$ to the fixed-point values $c_B^* = 0$ and $c_B^* = \infty$. By analogy with before, we denote the resulting scaling functions by $\Psi_B$, $\Xi_B$, $\Psi_A$, and $\Xi_A$, respectively.

We could perform the Fourier inversion only numerically. The numerical results for $\Psi_A(\mathsf{p}=0,\mathsf{t})$ and $\Psi_B(\mathsf{p}=0,\mathsf{t})$ one obtains with $\epsilon$ set equal to one are plotted in Fig. 3. Since at criticality, the $\mathbf{p} = \mathbf{0}$ susceptibility $\chi_{11}$ has a long-time tail $\sim t^{-(\zeta-1+\eta_\parallel^{\rm sp})/\zeta}$, the limiting values $\Xi_A^0 \equiv \Xi_A(0,0)$ and $\Xi_B^0 \equiv \Xi_B(0,0)$ both exists, which implies the small-$\mathsf{t}$ behavior

$$\Psi_{A,\,B}(\mathsf{p}=0,\mathsf{t}\to 0) \approx \Xi_{A,\,B}^0\,\mathsf{t}^{-(\zeta-1+\eta_\parallel^{\rm sp})/\zeta}\ . \tag{57}$$

Hence the difference between the dynamic surface universality classes manifest itself here only through different values of the universal amplitudes $\Xi_A^0$ and $\Xi_B^0$.

The asymptotic forms of $\Psi_{A,\,B}(0,\mathsf{t})$ in the *large*-$\mathsf{t}$ limit reflect the time-dependence of $\chi_{11}(\mathbf{p}=\mathbf{0},t,\tau>0,\tilde{c})$ for $\tilde{c} > 0$ and $\tilde{c} = 0$. Since the order parameter is a conserved density away from the surface, one anticipates a hydrodynamic long-time tail even in the case of model $B_A$. The asymptotic small-frequency dependencies given in (54) and (55) suggest distinct algebraic decays $\sim t^{-5/2}$ and $\sim t^{-3/2}$ for models $B_A$ and $B_B$, respectively.

In Appendix F the large-$\mathsf{t}$ behavior of the zero-loop scaling functions $\Psi_{A,B}^{\epsilon=0}(\mathsf{p},\mathsf{t})$ is worked out in detail. We find the asymptotic forms

$$\Psi_A^{\epsilon=0}(\mathsf{p},\mathsf{t}) = \frac{\sqrt{1+2\mathsf{p}^2}}{2\sqrt{\pi}\,(1+\mathsf{p}^2)^2}\frac{e^{-\mathsf{p}^2(1+\mathsf{p}^2)\mathsf{t}}}{\mathsf{t}^{3/2}}\left[\mathsf{p}^2 - \frac{3(5\mathsf{p}^2+5\mathsf{p}^6-2)}{4\,(1+\mathsf{p}^2)^2\,(1+2\mathsf{p}^2)^2}\,\frac{1}{\mathsf{t}} + O\left(\frac{1}{\mathsf{t}^2}\right)\right] \tag{58}$$

and

$$\Psi_B^{\epsilon=0}(\mathsf{p},\mathsf{t}) = \frac{\sqrt{1+2\mathsf{p}^2}}{\sqrt{\pi}}\frac{e^{-\mathsf{p}^2(1+\mathsf{p}^2)\mathsf{t}}}{\mathsf{t}^{1/2}}\left[\mathsf{p}^2 + \frac{2+\mathsf{p}^2}{4\,(1+2\mathsf{p}^2)^2}\,\frac{1}{\mathsf{t}} + O\left(\frac{1}{\mathsf{t}^2}\right)\right]. \tag{59}$$

For $\mathsf{p} > 0$, the results exhibit the expected exponential decay with the same relaxation rate $\mathsf{p}^2(1+\mathsf{p}^2)$. In the hydrodynamic limit $\mathsf{p} \to 0$, the relaxation rate vanishes, and we get the algebraic decay laws

$$\Psi_A(\mathsf{p}=0, \mathsf{t}\to\infty) \sim \mathsf{t}^{-5/2} \tag{60}$$

and

$$\Psi_B(\mathsf{p}=0, \mathsf{t}\to\infty) \sim \mathsf{t}^{-3/2} \ . \tag{61}$$

Since the above large-$\mathsf{t}$ behavior is of a noncritical, purely hydrodynamic, origin, its form should already be correctly obtained at zero-loop order. Terms beyond this order will affect the (suppressed) universal amplitudes involved in (60) and (61), but should neither change the exponents in these equations nor the exponential decay on a time scale $\propto \mathsf{p}^{-2}$ predicted in (58) and (59). Our one-loop results shown in Fig. 3 are in conformity with these expectations.

## V. CONCLUDING REMARKS

Let us conclude by summarizing the principal results of the present work.

Using field-theoretic RG methods, we investigated the dynamic surface critical behavior of the semi-infinite models $B_A$ and $B_B$ introduced in I and Ref. [16]. These models represent the dynamic surface universality classes of surface-bounded macroscopic systems whose dynamic bulk critical behavior is described by the familiar model B of Ref. [18]). They differ by the presence or absence of nonconservative surface terms.

Our first goal was to check whether the dynamic field theories of these models can indeed be renormalized in the manner asserted in I. To this end, we presented a variety of two-loop calculations through which we confirmed the claim of I that all renormalization factors involved are given by the known renormalization factors of the static theory. The resulting RG equation then lead to the conclusion that the critical exponents of the dynamic theory are expressible in terms of static (bulk and surface) critical indices. Since models $B_A$ and $B_B$ have the same thermodynamic equilibrium state, the dynamic critical exponents of both models are the same and hence do not differentiate between the respective two distinct dynamic surface universality classes that are associated with the same static surface universality class.

Our second goal was to investigate a characteristic dynamic quantity that is sensitive to the differences between models $B_A$ and $B_B$, and hence distinguishes between the corresponding dynamic surface universality classes. Using RG-improved perturbation theory, we computed the dynamic surface susceptibility $\hat{\chi}_{11}(\mathbf{p},\omega)$ for the static universality class of the special transition ($c = 0$) and the associated scaling functions $\hat{\Psi}_A$ and $\hat{\Psi}_B$. The $\epsilon$ expansions of the latter functions are given in (45) and (46). Their analogs $\Psi_A(0,\mathsf{t})$ and $\Psi_B(0,\mathsf{t})$ for $\chi_{11}(\mathbf{p}=\mathbf{0},t)$ are depicted in Fig. 3. They are clearly different; in particular, their asymptotic behavior for large values of the scaling argument $\mathsf{t} \sim t\tau^{\nu\zeta}$ is markedly different.

As our calculations show, detailed studies of dynamic critical behavior at surfaces by means of field-theoretic RG methods and the $\epsilon$ expansion, including the calculation of scaling functions, are feasible, albeit technically rather demanding. Unfortunately, we are at present not aware of experimental studies of dynamic surface critical behavior of systems belonging to the bulk universality class of model $B$. Once such studies become available, the measured observables could be computed by the methods described above.

## VI. ACKNOWLEDGEMENTS

F. W. would like to thank the Studienstiftung des Deutschen Volkes for a fellowship. We are also grateful to the Deutsche Forschungsgemeinschaft for partial support through Sonderforschungsbereich 237.

## APPENDIX A: $\epsilon$ EXPANSION OF DISTRIBUTIONS ASSOCIATED WITH AMPUTATED TWO-POINT GRAPHS

In the following, $\tau$ and $c$ are set to their (multi-)critical values $\tau = c = 0$. The response propagator $G$ then becomes

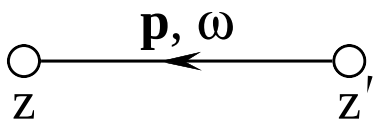

$$= \frac{1}{2\lambda_0(\kappa_+^2 - \kappa_-^2)} \left\{ -\frac{1}{\kappa_+} e^{-\kappa_+|z-z'|} - \frac{2\kappa_+\kappa_- + \tilde{c}_0\,(\kappa_+ - \kappa_-)}{\kappa_+[2\kappa_+\kappa_- + \tilde{c}_0\,(\kappa_+ + \kappa_-)]}\, e^{-\kappa_+(z+z')} \right.$$

$$+\frac{1}{\kappa_-}\,e^{-\kappa_-|z-z'|}+\frac{2\kappa_+\kappa_-+\tilde{c}_0\,(\kappa_--\kappa_+)}{\kappa_-[2\kappa_+\kappa_-+\tilde{c}_0\,(\kappa_++\kappa_-)]}\,e^{-\kappa_-(z+z')}$$

$$+\frac{2\tilde{c}_0}{2\kappa_+\kappa_-+\tilde{c}_0\,(\kappa_++\kappa_-)}\Big[e^{-(\kappa_+z+\kappa_-z')}-e^{-(\kappa_-z+\kappa_+z')}\Big]\Big\} \tag{A1}$$

with

$$\kappa_\pm=\Big[\mathbf{p}^2\pm\sqrt{i(\omega/\lambda_0)}\Big]^{1/2}\;. \tag{A2}$$

We also need the well-known static propagator

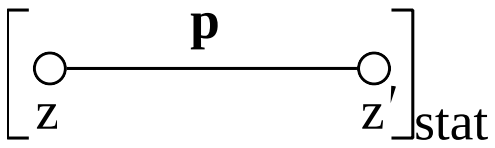

$$=\frac{1}{2p}\Big[e^{-p(z+z')}+e^{-p|z-z'|}\Big]\;. \tag{A3}$$

The required one-loop graph is proportional to its static counterpart. We have

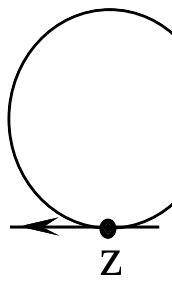

$$=\lambda_0\left[-\frac{u_0}{2}\,C(\mathbf{x},t;\mathbf{x},t)\right]$$

$$=\lambda_0$$

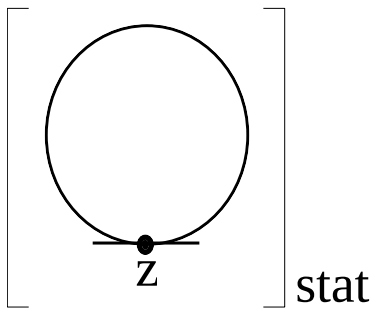

(A4)

$$=\;-\,\frac{\lambda_0u_0s_d}{2}\,\Gamma(1-\epsilon/2)\,z^{-2+\epsilon} \tag{A5}$$

$$=\frac{\lambda_0u_0s_d}{2}\left[\left(\frac{1}{\epsilon}+\frac{C_E}{2}\right)\delta'(z)-z_+^{-2}+O(\epsilon)\right]. \tag{A6}$$

Here $C_E$ denotes Euler's number, and $z_+^{-m+\epsilon}$ is a distribution whose definition for $\epsilon \gtrsim 0$ can be found in Ref. [22] or in the appendix of Ref. [2].

In a similar fashion we get

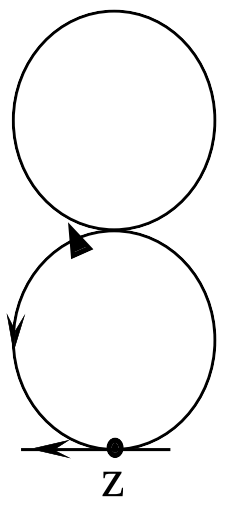

$$=\lambda_0$$

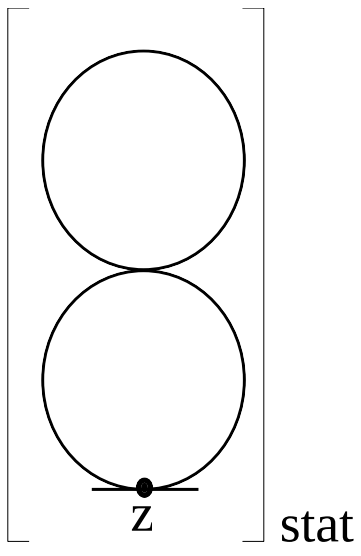

$$=\frac{\lambda_0(s_du_0)^2}{4}\left\{-\frac{1}{\epsilon^2}\,\delta'(z)+\frac{1}{\epsilon}\left[2z_+^{-2}+\frac{1-2C_E}{2}\,\delta'(z)\right]+O(\epsilon^0)\right\}. \tag{A7}$$

The graph that contains the static one-loop graph (A4) twice is not simply proportional to a static one. The $\epsilon$ expansion of this distribution can be determined by computing its action on the test function $e^{-\alpha z+\beta z'}$ (cf. Sec. III.A). This yields

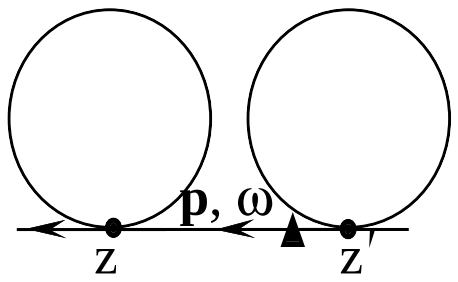

$$= \frac{\lambda_0(u_0 s_d)^2}{8}\Bigg\{\frac{1}{\epsilon^2}\Big[\delta(z)\,\delta'(z') + \delta'(z)\,\delta(z') + \frac{2(\kappa_+ + \kappa_-) + 4\tilde{c}_0}{2\kappa_+\kappa_- + \tilde{c}_0(\kappa_+ + \kappa_-)}\,\delta'(z)\,\delta'(z')\Big]$$
$$+ \frac{1}{\epsilon}\Big[(C_E - 1)[\delta(z)\delta'(z') + (z \leftrightarrow z')] + C_E\,\frac{2(\kappa_+ + \kappa_-) + 4\tilde{c}_0}{2\kappa_+\kappa_- + \tilde{c}_0(\kappa_+ + \kappa_-)}\,\delta'(z)\,\delta'(z')$$
$$- \frac{2}{2\kappa_+\kappa_- + \tilde{c}_0(\kappa_+ + \kappa_-)}\Big\{\Big[(\kappa_- + \tilde{c}_0)\,e^{-\kappa_+ z} + (\kappa_+ + \tilde{c}_0)\,e^{-\kappa_- z}\Big] z_+^{-2}\,\delta'(z')$$
$$+ (z \leftrightarrow z')\Big\}\Big] + O(\epsilon^0)\Bigg\}. \qquad \text{(A8)}$$

The Laurent part of the $\epsilon$ expansion of the remaining two-loop graph may be conveniently determined by expanding in powers of $\omega$. It is easy to see that the $\omega$-dependent part of the graph is regular in $\epsilon$. (Use power counting and the form of the divergent subgraphs.) Hence the poles result from the $\omega = 0$ term, which is proportional to the corresponding graph of the static theory. It follows that

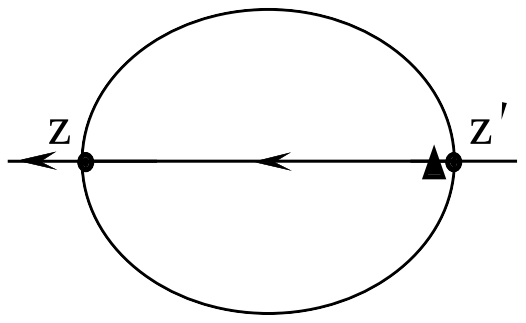

$=$

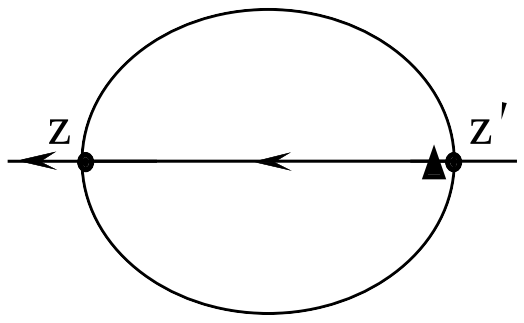

$\Big|_{\omega=0} + \omega\Big[\frac{d}{d\omega}$

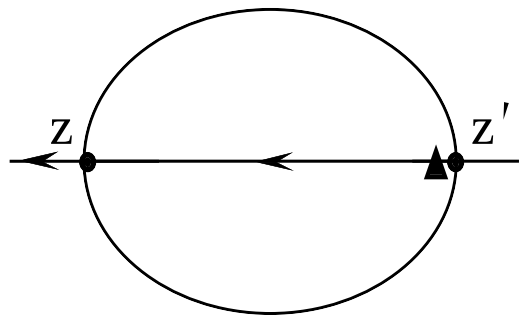

$\Big]_{\omega=0} + O(\omega^2)$

$= \lambda_0 \Big[$

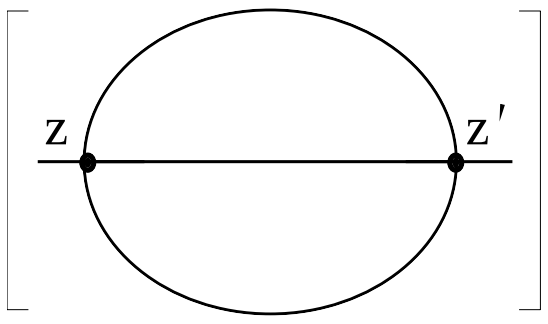

$\Big]_{\text{stat}} + O(\epsilon^0)$ (A9)

$$= \lambda_0(s_d\,u_0)^2\Bigg\{ - \frac{1}{2\epsilon^2}\Big[\delta(z)\,\delta'(z') + \delta'(z)\,\delta(z')\Big]$$
$$+ \frac{1}{4\epsilon}\Big(\frac{1}{3}\Big[\delta''(z - z') - p^2\delta(z - z')\Big] + 4z_+^{-2}\,\delta(z - z')$$
$$- \Big[\frac{9}{4} + 2C_E\Big]\Big[\delta(z)\,\delta'(z') + \delta'(z)\,\delta(z')\Big]\Big) + O(\epsilon^0)\Bigg\}\,. \qquad \text{(A10)}$$

## APPENDIX B: GRAPHS OF $G^{(0,1;1,0)}$ AND $W^{(0,1;1,0)}$ FOR $\tau = c = 0$

Utilizing the results of Appendix A, the $\epsilon$ expansion of the desired graphs of $G^{(0,1;1,0)}$ and $W^{(0,1;1,0)}$ for $\tau = c = 0$ can be derived in a straightforward fashion. We first list our results for the graphs of $G^{(0,1;1,0)}$. To one-loop order we have the contributions

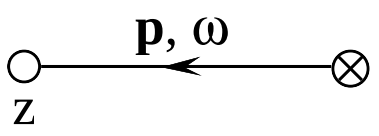

$$= \frac{2\left[\kappa_+ e^{-\kappa_+ z} - \kappa_-\, e^{-\kappa_- z}\right]}{\lambda_0\left(\kappa_+^2 - \kappa_-^2\right)\left[2\kappa_+\kappa_- + \tilde{c}_0\left(\kappa_+ + \kappa_-\right)\right]} \tag{B1}$$

and

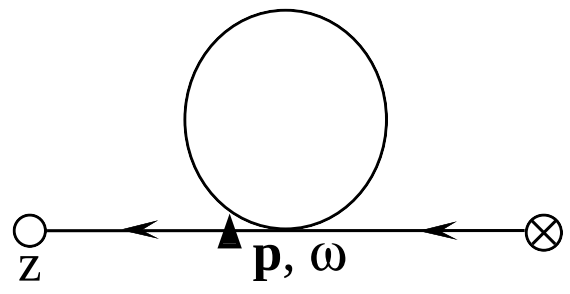

$$= -\,\frac{s_d\, u_0\left[E_+\, e^{-\kappa_+ z} - E_-\, e^{-\kappa_- z}\right]}{4\lambda_0\left(\kappa_+^2 - \kappa_-^2\right)\left[2\kappa_+\kappa_- + \tilde{c}_0\left(\kappa_+ + \kappa_-\right)\right]} + O(\epsilon) \tag{B2}$$

with

$$\begin{aligned} E_\pm(z) = &-2\kappa_\mp\, e^{2\kappa_\pm z}\,\mathrm{Ei}(-2\kappa_\pm z) + (\kappa_\pm + \kappa_\mp)\, e^{2\kappa_\pm z}\,\mathrm{Ei}[-(\kappa_\pm + \kappa_\mp) z] \\ &+ (\kappa_\mp - \kappa_\pm)\left\{\ln(\kappa_\mp - \kappa_\pm) - \mathrm{Ei}[-(\kappa_\mp - \kappa_\pm) z]\right\} \\ &+ \frac{1}{2\kappa_\pm\kappa_\mp + \tilde{c}_0(\kappa_\pm + \kappa_\mp)}\Big\{2\tilde{c}_0(C_E - 1)(\kappa_\pm^2 - \kappa_\mp^2) + 4\tilde{c}_0\kappa_\pm\kappa_\mp\ln(2\kappa_\mp) \\ &+ \Big[2\kappa_\pm\kappa_\mp + \tilde{c}_0(\kappa_\pm - \kappa_\mp)\Big]\Big[-2\kappa_\mp\ln(2\kappa_\pm) + (\kappa_\pm + \kappa_\mp)\ln(\kappa_\pm + \kappa_\mp)\Big] \\ &- 2\tilde{c}_0\kappa_\mp(\kappa_\pm + \kappa_\mp)\ln(\kappa_\pm + \kappa_\mp)\Big\}\,, \end{aligned} \tag{B3}$$

where Ei is the exponential integral function.

Our results for the two-loop graphs are

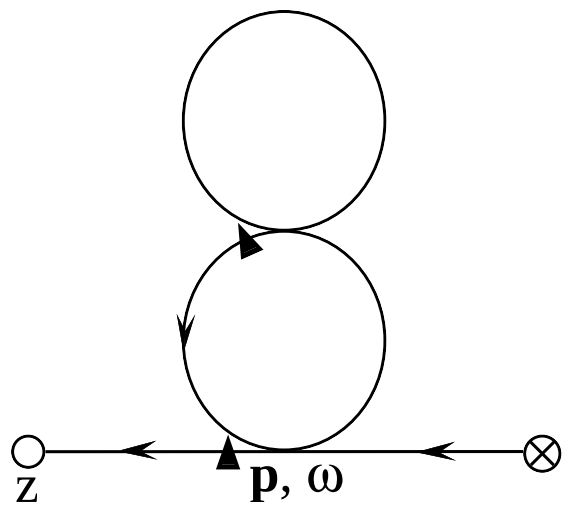

$$= \frac{(s_d\, u_0)^2\left[E_+\, e^{-\kappa_+ z} - E_-\, e^{-\kappa_- z}\right]}{4\lambda_0(\kappa_+^2 - \kappa_-^2)[2\kappa_+\kappa_- + \tilde{c}_0(\kappa_+ + \kappa_-)]}\,\frac{1}{\epsilon} + O(\epsilon^0)\,, \tag{B4}$$

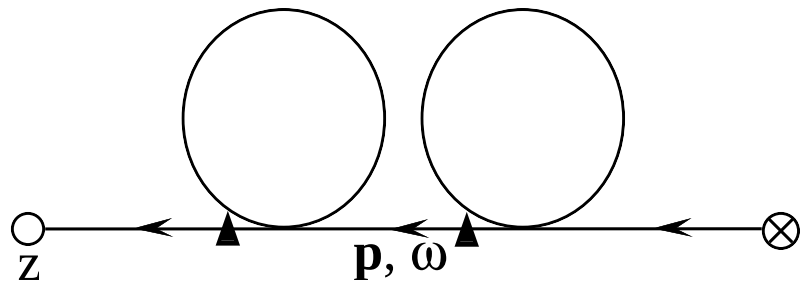

$$= O(\epsilon^0)\,, \tag{B5}$$

and

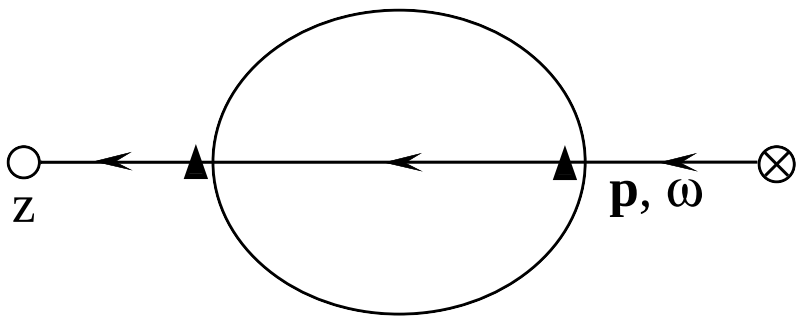

$$= -\frac{(s_d\, u_0)^2 \left[G_+\, e^{-\kappa_+ z} - G_-\, e^{-\kappa_- z}\right]}{48\lambda_0(\kappa_+^2 - \kappa_-^2)\left[2\kappa_+\kappa_- + \tilde{c}_0(\kappa_+ + \kappa_-)\right]^2 \kappa_+\kappa_-}\,\frac{1}{\epsilon} + O(\epsilon^0) \tag{B6}$$

with

$$G_\pm(z) = 24\kappa_\pm\kappa_\mp\left[2\kappa_\pm\kappa_\mp + \tilde{c}_0(\kappa_\pm + \kappa_\mp)\right]E_\pm(z) - 2z\kappa_\mp^3\kappa_\pm(\kappa_\pm^2 - \kappa_\mp^2) + 2\kappa_\mp^3(\kappa_\mp^2 + 3\kappa_\pm^2)$$
$$+ \tilde{c}_0\left[-z\kappa_\mp^2(\kappa_\pm^2 - \kappa_\mp^2)(\kappa_\pm + \kappa_\mp) + 4\kappa_\pm\kappa_\mp^2(\kappa_\pm + \kappa_\mp) - \kappa_\pm^4 + \kappa_\mp^4\right]. \tag{B7}$$

For the graphs of $W^{(0,1;1,0)}$ we find

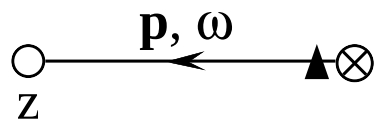

$$= \frac{(\kappa_- + \tilde{c}_0)\, e^{-\kappa_+ z} + (\kappa_+ + \tilde{c}_0)\, e^{-\kappa_- z}}{\lambda_0\left[2\kappa_+\kappa_- + \tilde{c}_0(\kappa_+ + \kappa_-)\right]}\;, \tag{B8}$$

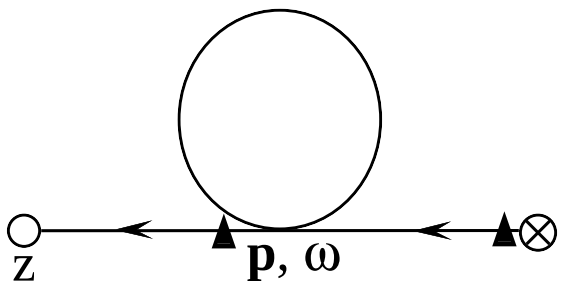

$$= \frac{s_d\, u_0}{2\lambda_0\left[2\kappa_+\kappa_- + \tilde{c}_0(\kappa_+ + \kappa_-)\right]}\left\{\frac{1}{\epsilon}\left[(\kappa_- + \tilde{c}_0)e^{-\kappa_+ z} + (\kappa_+ + \tilde{c}_0)e^{-\kappa_- z}\right]\right.$$
$$\left. + \left(\left[\frac{C_E\,(\kappa_- + \tilde{c}_0)}{2} - \frac{F_+}{4}\right]e^{-\kappa_+ z} + \left[\frac{C_E(\kappa_+ + \tilde{c}_0)}{2} - \frac{F_-}{4}\right]e^{-\kappa_- z}\right) + O(\epsilon)\right\} \tag{B9}$$

with

$$F_\pm(z) = 2(\kappa_\mp + \tilde{c}_0)e^{2\kappa_\pm z}\mathrm{Ei}(-2\kappa_\pm z) + \frac{(\kappa_\pm + \tilde{c}_0)(\kappa_\pm + \kappa_\mp)}{\kappa_\pm}\, e^{2\kappa_\pm z}\mathrm{Ei}[-(\kappa_\pm + \kappa_\mp)z]$$
$$+ \frac{(\kappa_\pm + \tilde{c}_0)(\kappa_\mp - \kappa_\pm)}{\kappa_\pm}\Big\{\ln(\kappa_\mp - \kappa_\pm) - \mathrm{Ei}\left[-(\kappa_\mp - \kappa_\pm)z\right]\Big\} + 4(\kappa_\mp + \tilde{c}_0)(C_E - 1)$$
$$+ \frac{1}{\kappa_\pm[2\kappa_\pm\kappa_\mp + \tilde{c}_0(\kappa_\pm + \kappa_\mp)]}\Big\{2\kappa_\pm(\kappa_\mp + \tilde{c}_0)[2\kappa_\pm\kappa_\mp + \tilde{c}_0(\kappa_\pm - \kappa_\mp)]\ln(2\kappa_\pm)$$
$$+ 4\tilde{c}_0\kappa_\pm\kappa_\mp(\kappa_\pm + \tilde{c}_0)\ln(2\kappa_\mp) + \Big[2\kappa_\pm^2\kappa_\mp(\kappa_\pm + \kappa_\mp)$$
$$+ \tilde{c}_0\kappa_\pm(\kappa_\pm^2 + 4\kappa_\mp\kappa_\pm + 3\kappa_\mp^2) + \tilde{c}_0^2(3\kappa_\pm^2 + 2\kappa_\pm\kappa_\mp - \kappa_\mp^2)\Big]\ln(\kappa_\pm + \kappa_\mp)\Big\}\;, \tag{B10}$$

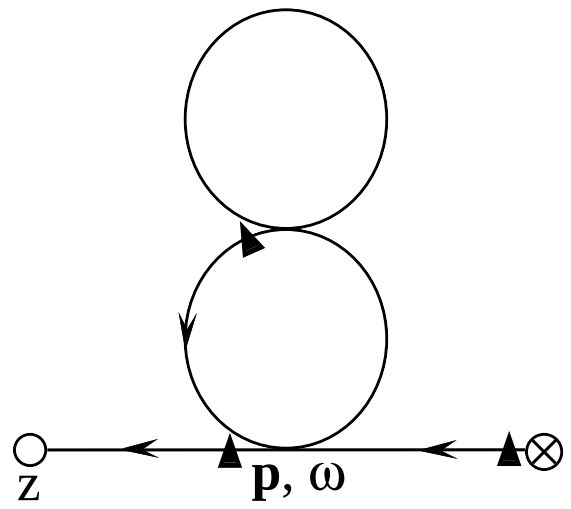

$$= \frac{(s_d\, u_0)^2}{4\lambda_0[2\kappa_+\kappa_- + \tilde{c}_0(\kappa_+ + \kappa_-)]}\Bigg\{ -\frac{1}{\epsilon^2}\Big[(\kappa_- + \tilde{c}_0)e^{-\kappa_+ z} + (\kappa_+ + \tilde{c}_0)e^{-\kappa_- z}\Big]$$

$$+ \frac{1}{2\epsilon}\Big([F_+ - (\kappa_- + \tilde{c}_0)(2C_E - 1)]e^{-\kappa_+ z}$$
$$+ [F_- - (\kappa_+ + \tilde{c}_0)(2C_E - 1)]e^{-\kappa_- z}\Big) + O(\epsilon^0)\Bigg\}\ , \qquad \text{(B11)}$$

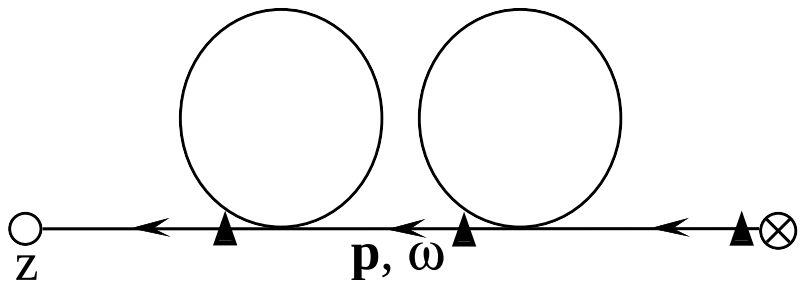

$$= \frac{(s_d\, u_0)^2}{8\lambda_0[2\kappa_+\kappa_- + \tilde{c}_0(\kappa_+ + \kappa_-)]}\Bigg\{\frac{1}{\epsilon^2}\Big[(\kappa_- + \tilde{c}_0)e^{-\kappa_+ z}$$

$$+ (\kappa_+ + \tilde{c}_0)e^{-\kappa_- z}\Big] + \frac{1}{\epsilon}\Big(\Big[(\kappa_- + \tilde{c}_0)(C_E - 1) - \tfrac{1}{2}F_+\Big]e^{-\kappa_+ z}$$
$$+ \Big[(\kappa_+ + \tilde{c}_0)(C_E - 1) - \tfrac{1}{2}F_-\Big]e^{-\kappa_- z}\Big) + O(\epsilon^0)\Bigg\}, \qquad \text{(B12)}$$

and

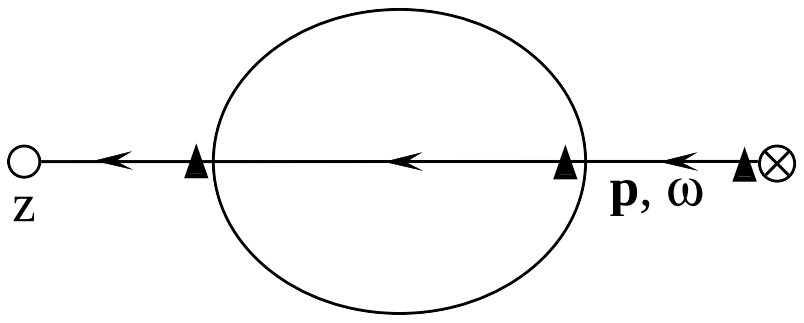

$$= \frac{(s_d\, u_0)^2}{\lambda_0[2\kappa_+\kappa_- + \tilde{c}_0(\kappa_+ + \kappa_-)]}\Bigg\{ -\frac{1}{2\epsilon^2}\Big[(\kappa_- + \tilde{c}_0)e^{-\kappa_+ z}$$

$$+ (\kappa_+ + \tilde{c}_0)e^{-\kappa_- z}\Big] + \frac{1}{\epsilon}\,\frac{H_+\, e^{-\kappa_+ z} + H_-\, e^{-\kappa_- z}}{96\kappa_+\kappa_-[2\kappa_+\kappa_- + \tilde{c}_0(\kappa_+ + \kappa_-)]}$$
$$+ O(\epsilon^0)\Bigg\} \qquad \text{(B13)}$$

with

$$H_\pm(z) = \Big[2\kappa_\mp^3\kappa_\pm z\,(\kappa_\pm^2 - \kappa_\mp^2) + 48F_\pm(z)\,\kappa_\pm^2\kappa_\mp^2 - 2\kappa_\mp^3(55\kappa_\pm^2 + \kappa_\mp^2) - 96C_E\kappa_\pm^2\kappa_\mp^3\Big]$$
$$+ \tilde{c}_0\Big\{\Big[z\kappa_\mp^2(\kappa_\pm^2 - \kappa_\mp^2) - 48C_E\kappa_\mp^2\kappa_\pm\Big](3\kappa_\pm + \kappa_\mp) + 24\kappa_\pm\kappa_\mp(\kappa_\pm + \kappa_\mp)\,F_\pm(z)$$
$$- \kappa_\pm^4 - 164\kappa_\mp^2\kappa_\pm^2 - 56\kappa_\mp^3\kappa_\pm - 3\kappa_\mp^4\Big\} + \tilde{c}_0^2\Big\{-\kappa_\pm^3 - \kappa_\mp^3$$
$$+ \Big[z\kappa_\mp(\kappa_\pm^2 - \kappa_\mp^2) - (48C_E + 55)\kappa_\pm\kappa_\mp\Big](\kappa_\pm + \kappa_\mp)\Big\}. \qquad \text{(B14)}$$

## APPENDIX C: REPARAMETRIZED GRAPHS OF $G_R^{(0,1;1,0)}$ AND $W_R^{(0,1;1,0)}$

In the foregoing appendix the graphs of the bare functions $G_R^{(0,1;1,0)}$ and $W_R^{(0,1;1,0)}$ were obtained. Here we express these results in terms of the renormalized variables $u$ and $\lambda$, employing the reparametrizations (13b) and (14a). For notational simplicity we set $\mu = 1$. We need the $n = 1$ relations

$$\lambda_0 = \lambda \left[1 - \frac{1}{12\epsilon} u^2 + O(u^3)\right] , \quad u_0 s_d = u \left[1 + \frac{3}{\epsilon} u + O(u^2)\right] , \tag{C1}$$

and

$$\kappa_\pm = k_\pm \left[1 + \left(1 - \frac{k_\mp^2}{k_\pm^2}\right) \frac{u^2}{96\epsilon} + O(u^3)\right] \tag{C2}$$

with

$$k_\pm = \sqrt{\mathbf{p}^2 \pm \sqrt{i\omega/\lambda}} \, . \tag{C3}$$

Since we restrict ourselves to two-loop order, we must retain only terms to order $u^2$. Hence for the two-loop graphs of Appendix B, the reparametrization amounts to the substitutions $u_0 s_d \to u$, $\lambda_0 \to \lambda$, and $\tilde{c}_0 \to \tilde{c}$. Accordingly we list here only the results for the zero-loop and one-loop graphs. To indicate that the graphs are expressed in terms of renormalized parameters, we use the notation $[.]_{\mathrm{Rep}}$.

Upon introducing the quantities

$$\begin{aligned} \mathcal{L}_\pm(z) \equiv & -2k_\mp^3(k_\mp^2 + 3k_\pm^2) + 2k_\pm k_\mp^3 z (k_\pm^2 - k_\mp^2) \\ & + \tilde{c}\Big[z k_\mp^2 (k_\pm^2 - k_\mp^2)(k_\pm + k_\mp) - 4k_\mp^2 k_\pm (k_\pm + k_\mp) + k_\pm^4 - k_\mp^4\Big] \end{aligned} \tag{C4}$$

and

$$\mathcal{E}_\pm \equiv E_\pm|_{\kappa_\pm \to k_\pm,\, \tilde{c}_0 \to \tilde{c}} \, , \tag{C5}$$

our results for the graphs of $G_R^{(0,1;1,0)}$ may be written as

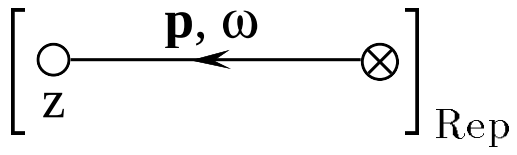

$$= \frac{1}{\lambda(k_+^2 - k_-^2)[2k_+k_- + \tilde{c}(k_+ + k_-)]} \Big\{ 2k_+ \, e^{-k_- z} - 2k_- \, e^{-k_+ z} + \frac{u^2}{\epsilon} \frac{\mathcal{L}_+ e^{-k_+ z} + \mathcal{L}_- e^{-k_- z}}{48 k_+ k_- [2k_+k_- + \tilde{c}(k_+ + k_-)]} + O(u^3) \Big\} \tag{C6}$$

and

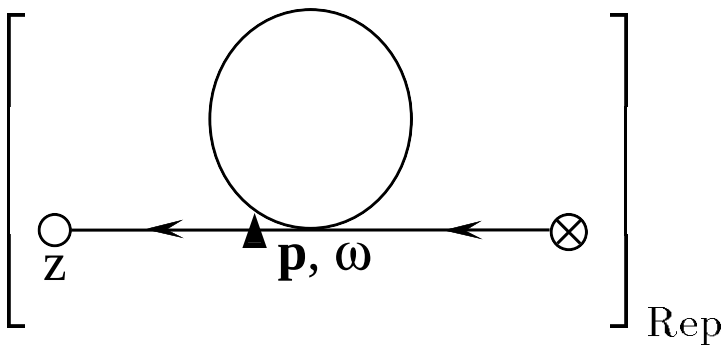

$$= -\frac{u}{4\lambda(k_+^2 - k_-^2)[2k_+k_- + \tilde{c}(k_+ + k_-)]} \Big\{ \mathcal{E}_+ e^{-k_+ z} - \mathcal{E}_- e^{-k_- z} + O(\epsilon) + \frac{3u}{\epsilon} \Big[ \mathcal{E}_+ e^{-k_+ z} - \mathcal{E}_- e^{-k_- z} + O(\epsilon) \Big] + O(u^3) \Big\}. \tag{C7}$$

Our results for the graphs of $W_R^{(0,1;1,0)}$ are

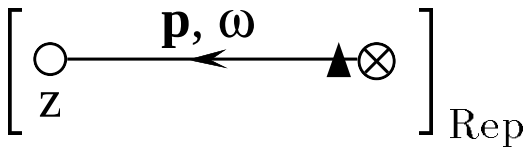

$$= \frac{1}{\lambda[2k_+k_- + \tilde{c}(k_+ + k_-)]} \Big\{ (k_- + \tilde{c}) e^{-k_+ z} + (k_+ + \tilde{c}) e^{-k_- z} - \frac{u}{2\epsilon} \Big[ (k_- + \tilde{c}) e^{-k_+ z} + (k_+ + \tilde{c}) e^{-k_- z} + O(\epsilon^0) \Big] + u^2 \Big[ -\frac{5}{8\epsilon^2} \Big\{ (k_- + \tilde{c}) e^{-k_+ z} + (k_+ + \tilde{c}) e^{-k_- z} \Big\} + \frac{1}{\epsilon} \frac{\mathcal{M}_+ e^{-k_+ z} + \mathcal{M}_- e^{-k_- z}}{96 k_+ k_- [2k_+k_- + \tilde{c}(k_+ + k_-)]} + O(\epsilon^0) \Big] + O(u^3) \Big\} \tag{C8}$$

with

$$\mathcal{M}_\pm = 2k_\mp^3(k_\mp^2 + 55k_\pm^2) + 2z k_\pm k_\mp^3 (k_\mp^2 - k_\pm^2) + \tilde{c} \Big[ k_\pm^4 + 164 k_\mp^2 k_\pm^2 + 56 k_\pm k_\mp^3 + 3k_\mp^4 + z k_\mp^2 (k_\mp^2 - k_\pm^2)(3k_\pm + k_\mp) \Big] + \tilde{c}^2 \Big[ k_\pm^3 + k_\mp^3 + 55 k_\pm k_\mp (k_\pm + k_\mp) + z k_\mp (k_\mp^2 - k_\pm^2)(k_\pm + k_\mp) \Big] \tag{C9}$$

and

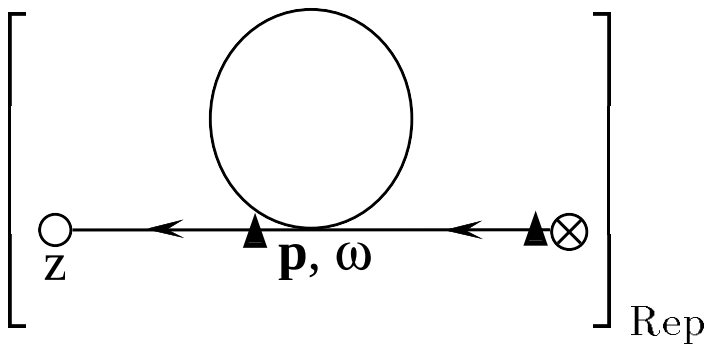

$$= \frac{u}{2\lambda[2k_+k_- + \tilde{c}(k_+ + k_-)]} \Big\{ \frac{1}{\epsilon} \Big[ (k_- + \tilde{c}) e^{-k_+ z} + (k_+ + \tilde{c}) e^{-k_- z} \Big] + \Big[ \tfrac{1}{2} C_E \, (k_- + \tilde{c}) - \tfrac{1}{4} \mathcal{F}_+ \Big] e^{-k_+ z} + \Big[ \tfrac{1}{2} C_E \, (k_+ + \tilde{c}) - \tfrac{1}{4} \mathcal{F}_- \Big] e^{-k_- z} + O(\epsilon) + \frac{5u}{2\epsilon^2} \Big( (k_- + \tilde{c}) e^{-k_+ z} + (k_+ + \tilde{c}) e^{-k_- z} + \epsilon \Big[ \tfrac{1}{2} C_E \, (k_- + \tilde{c}) - \tfrac{1}{4} \mathcal{F}_+ \Big] e^{-k_+ z} + \epsilon \Big[ \tfrac{1}{2} C_E \, (k_+ + \tilde{c}) - \tfrac{1}{4} \mathcal{F}_- \Big] e^{-k_- z} + O(\epsilon^0) \Big) + O(u^3) \Big\} \tag{C10}$$

with

$$\mathcal{F}_\pm \equiv F_\pm\big|_{\kappa_\pm \to k_\pm,\, \tilde{c}_0 \to \tilde{c}} \;. \tag{C11}$$

## APPENDIX D: AMPUTATED GRAPHS WITH AN INSERTION OF$(\phi\Delta\tilde{\phi})_s$ AND GRAPHS OF $G^{(0,1;1,0;1)}$

The results listed below were obtained by following our previous strategy: We first computed the Laurent expansion of the distributions associated with amputated graphs of $G^{(0,1;1,0;1)}$, relating them to expressions known from the static theory whenever possible. Subsequently we utilized these results together with those of Appendices A–C to compute the Laurent expansion of the graphs of $G^{(0,1;1,0;1)}$ to the required order in $\epsilon$.

The graphs of the static theory appearing in the formulae below involve an insertion of the surface operator $[\frac{1}{2}(\phi^2)_s]_{\mathbf{P}}$, the static analog of $-\lambda_0[(\phi\Delta\tilde{\phi})_s]_{\mathbf{P},\Omega}$. As discussed in the main text (cf. Sec. III.C), we have set $\mathbf{P} = \mathbf{0}$ and $\Omega = 0$.

It should also be recalled that we restrict ourselves to the case of model $B_B$, setting $\tilde{c}_0 = 0$ in this appendix. Accordingly the functions $F_\pm$ and $G_\pm$ appearing in the results listed below must be interpreted as the expressions to which (B10) and (B7) reduce for this particular choice of $\tilde{c}_0$.

Our results for the amputated graphs are

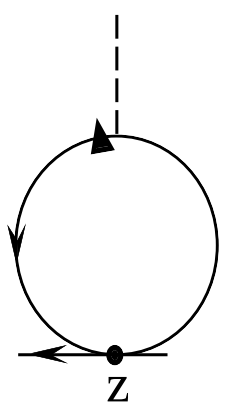

$= \lambda_0$

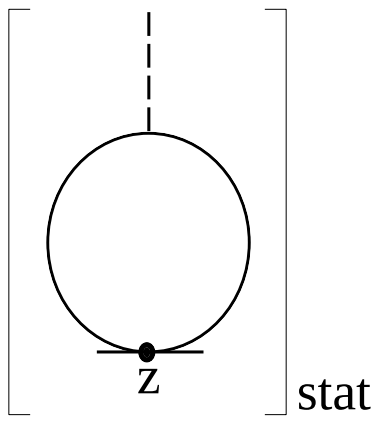

$$= -\lambda_0 s_d u_0 \left[\frac{2}{\epsilon}\delta(z) + (2 + C_E)\delta(z) + 2z_+^{-1} + O(\epsilon)\right], \tag{D1}$$

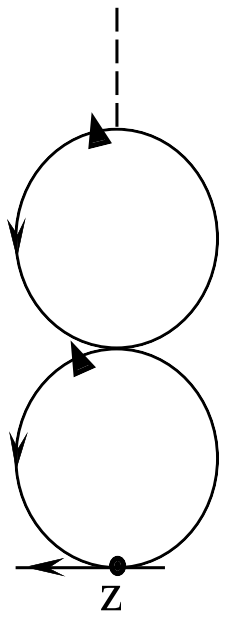

$$= \lambda_0 (s_d u_0)^2 \left\{ \frac{3}{\epsilon^2}\delta(z) + \frac{1}{\epsilon}\left[(7+3C_E)\delta(z) + 6z_+^{-1}\right] + O(\epsilon^0)\right\}, \tag{D2}$$

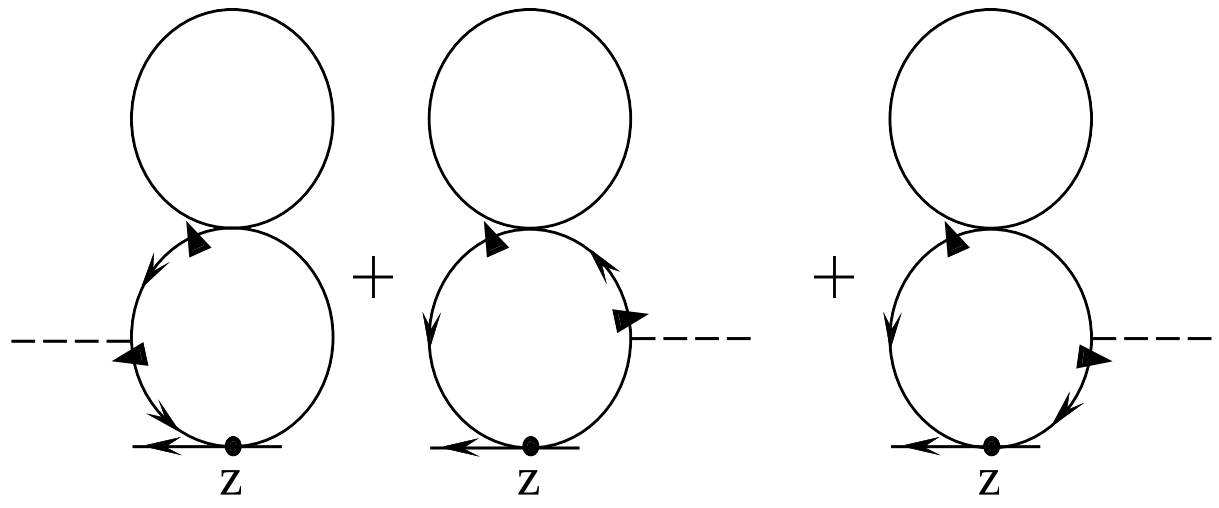

$= \lambda_0$

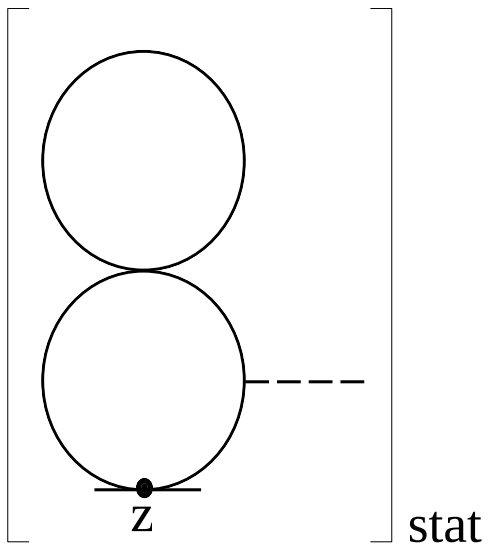

(D3)

$$= \lambda_0 (s_d u_0)^2 \left\{ -\frac{1}{\epsilon^2}\delta(z) - \frac{1}{\epsilon}\left[(3+C_E)\delta(z) + 2z_+^{-1}\right] + O(\epsilon^0)\right\}, \tag{D4}$$

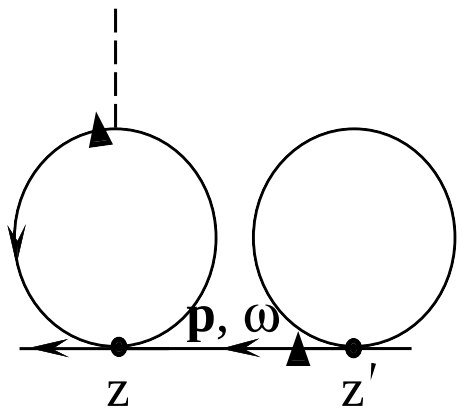

=

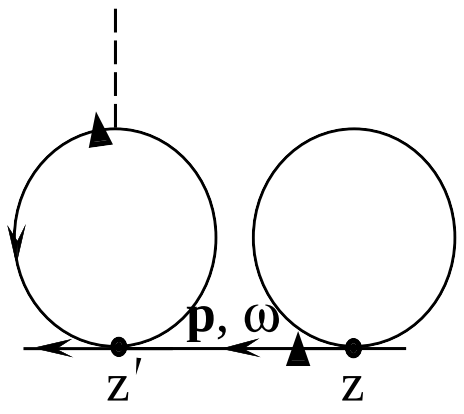

(D5)

$$\begin{aligned} &= -\frac{\lambda_0 (s_d u_0)^2}{2\epsilon^2}\Bigg\{ \frac{\kappa_+ + \kappa_-}{\kappa_+\kappa_-}\,\delta'(z')\delta(z) + \delta(z)\delta(z') \\ &\quad + \epsilon\Bigg[C_E\,\delta(z')\delta(z) + (1+C_E)\frac{\kappa_+ + \kappa_-}{\kappa_+\kappa_-}\,\delta'(z')\delta(z) \\ &\quad - {z'}_+^{-2}\delta(z)\Big(\kappa_+^{-1}e^{-\kappa_+ z'} + \kappa_-^{-1}e^{-\kappa_- z'}\Big) \\ &\quad + z_+^{-1}\delta'(z')\Big(\kappa_+^{-1}e^{-\kappa_+ z} + \kappa_-^{-1}e^{-\kappa_- z}\Big)\Bigg] + O(\epsilon^2)\Bigg\}\,, \end{aligned} \tag{D6}$$

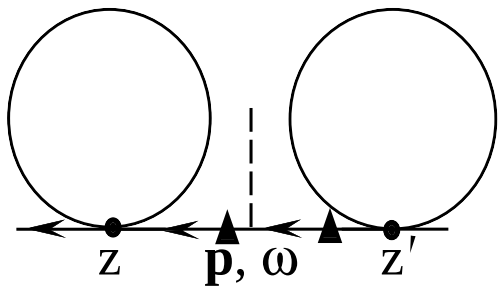

$$\begin{aligned} &= \frac{\lambda_0 (s_d u_0)^2}{16\epsilon^2}\Bigg(\left[\kappa_+^{-1} + \kappa_-^{-1}\right]^2\delta'(z)\delta'(z') + 4\delta(z)\delta(z') \\ &\quad + 2\left[\kappa_+^{-1} + \kappa_-^{-1}\right]\left[\delta'(z)\delta(z') + \delta(z)\delta'(z')\right] \\ &\quad + \epsilon\Bigg\{C_E\left[\kappa_+^{-1} + \kappa_-^{-1}\right]^2\delta'(z)\delta'(z') + 4C_E\,\delta(z)\delta(z') \end{aligned}$$

$$+ 2C_E \left[\kappa_+^{-1} + \kappa_-^{-1}\right] \left[\delta'(z)\delta(z') + \delta(z)\delta'(z')\right]$$
$$- z'^{-2}_+ \left[\left(\kappa_+^{-1} + \kappa_-^{-1}\right)\delta'(z) + 2\delta(z)\right] \left[\kappa_+^{-1} e^{-\kappa_+ z'} + \kappa_-^{-1} e^{-\kappa_- z'}\right]$$
$$- z_+^{-2} \left[\left(\kappa_+^{-1} + \kappa_-^{-1}\right)\delta'(z') + 2\delta(z')\right] \left[\kappa_+^{-1} e^{-\kappa_+ z} + \kappa_-^{-1} e^{-\kappa_- z}\right] \Bigg\}$$
$$+ O(\epsilon^2)\Bigg), \tag{D7}$$

and

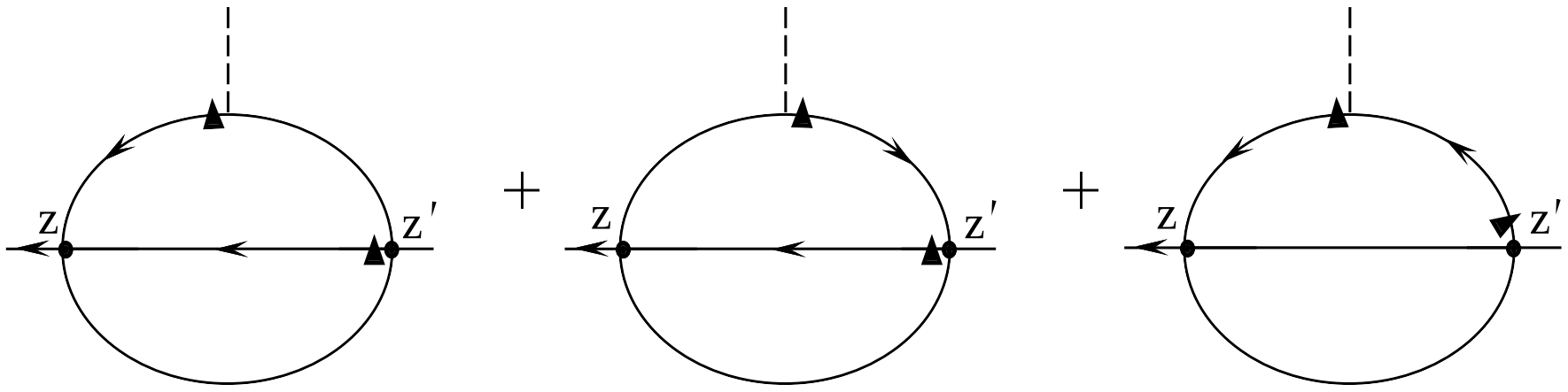

$= \lambda_0$

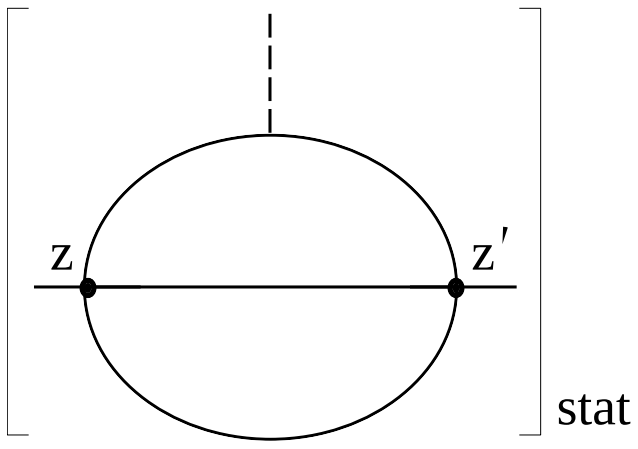

$$+ O(\epsilon^0) \tag{D8}$$

$$= \lambda_0 (s_d u_0)^2 \Big\{ \frac{2}{\epsilon^2}\,\delta(z)\delta(z') + \frac{1}{\epsilon}\Big[4 z_+^{-1}\delta(z - z') + \Big(2C_E + 5 + \tfrac{1}{3}\pi^2\Big)\delta(z)\delta(z')\Big] + O(\epsilon^0)\Big\}. \tag{D9}$$

We now turn to the graphs of $G^{(0,1;1,0;1)}$. In addition to the above results and those of Appendix A, we need the graph

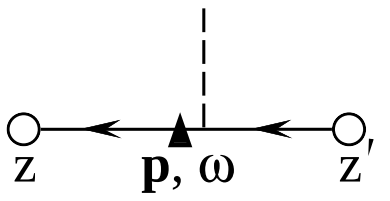

$= -\dfrac{d}{dc_0}\Big[$ 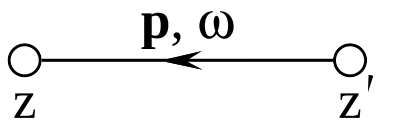 $\Big]_{c_0=0,\,\tilde{c}_0=0}$ (D10)

$$= \frac{1}{2\lambda_0(\kappa_+^2 - \kappa_-^2)} \left(\kappa_+^{-1} e^{-\kappa_+ z} + \kappa_-^{-1} e^{-\kappa_- z}\right) \left(\kappa_-^{-1} e^{-\kappa_- z'} - \kappa_+^{-1} e^{-\kappa_+ z'}\right). \tag{D11}$$

As an immediate consequence we have

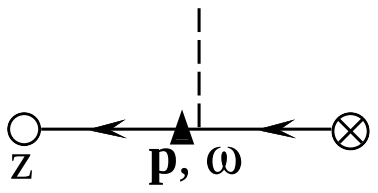

$$= \frac{1}{2\lambda_0(\kappa_+ + \kappa_-)\kappa_+\kappa_-} \left[\kappa_+^{-1} e^{-\kappa_+ z} + \kappa_-^{-1} e^{-\kappa_- z}\right]. \tag{D12}$$

For the one and two-loop graphs we find the results

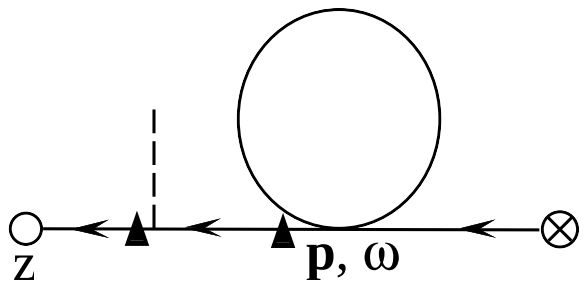

$= \dfrac{s_d u_0}{2}\Big\{\Big[\dfrac{1}{\epsilon} + \dfrac{C_E}{2}\Big]$ 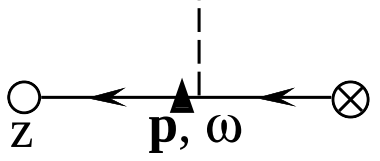 $- \dfrac{1}{2\lambda_0(\kappa_+^2 - \kappa_-^2)\kappa_+^2\kappa_-^2}$

$$\cdot \Big\{ \kappa_- e^{-\kappa_+ z} \Big[ (\kappa_+ - \kappa_-)(C_E - 1) - \kappa_- \ln 2\kappa_+ + \kappa_+ \ln 2\kappa_- \Big]$$
$$- \kappa_+ e^{-\kappa_- z} \Big[ (\kappa_- - \kappa_+)(C_E - 1) - \kappa_+ \ln 2\kappa_- + \kappa_- \ln 2\kappa_+ \Big] \Big\}$$
$$+ O(\epsilon) \Big\}, \tag{D13}$$

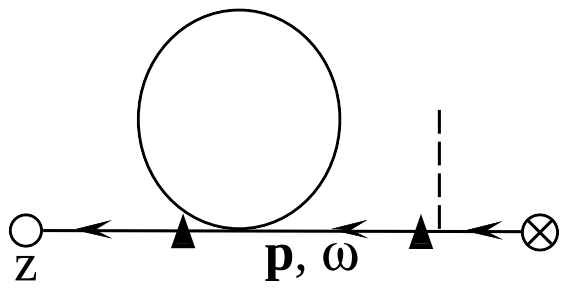

$$= \frac{s_d u_0}{2} \Big\{ \frac{1}{\epsilon}$$

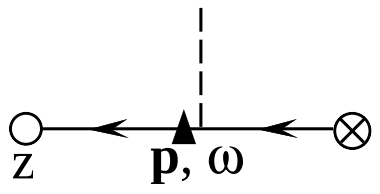

$$+ \frac{1}{2\lambda_0(\kappa_+ + \kappa_-)\kappa_+^2 \kappa_-^2}$$
$$\cdot \Big[ \Big( \tfrac{1}{2} C_E \kappa_- - \tfrac{1}{4} F_+ \Big) e^{-\kappa_+ z} + \Big( \tfrac{1}{2} C_E \kappa_+ - \tfrac{1}{4} F_- \Big) e^{-\kappa_- z} \Big] + O(\epsilon) \Big\}, \tag{D14}$$

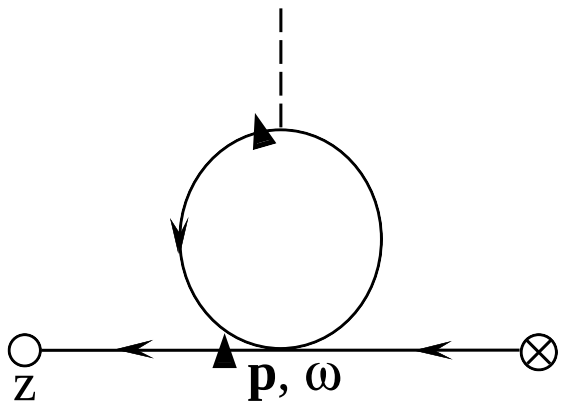

$$= s_d u_0 \Big\{ - \frac{2}{\epsilon}$$

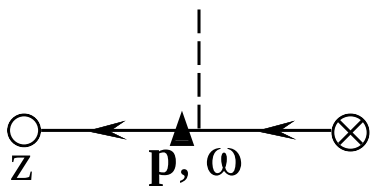

$$- \frac{1}{2\lambda_0(\kappa_+^2 - \kappa_-^2)\kappa_+^2 \kappa_-^2}$$
$$\cdot \Big( \Big[ (2 + C_E)(\kappa_+ - \kappa_-) + \kappa_+ G_+ \Big] \kappa_- e^{-\kappa_+ z}$$
$$- \Big[ (2 + C_E)(\kappa_- - \kappa_+) + \kappa_- G_- \Big] \kappa_+ e^{-\kappa_- z} \Big) + O(\epsilon) \Big\}, \tag{D15}$$

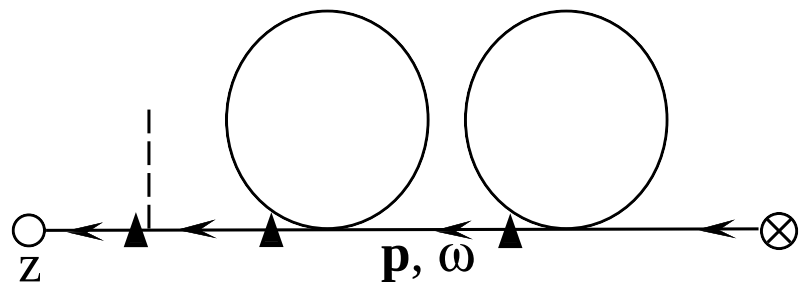

$$= \frac{(s_d u_0)^2}{8} \Big\{ \frac{1}{\epsilon^2} - \frac{1}{\epsilon} \Big[ C_E - 1 + \frac{2\kappa_- \ln 2\kappa_+}{\kappa_- - \kappa_+} + \frac{2\kappa_+ \ln 2\kappa_-}{\kappa_+ - \kappa_-} \Big]$$
$$+ O(\epsilon^0) \Big\}$$

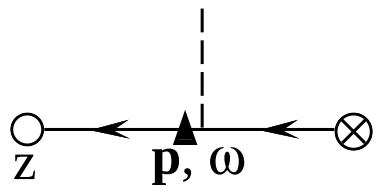

$$, \tag{D16}$$

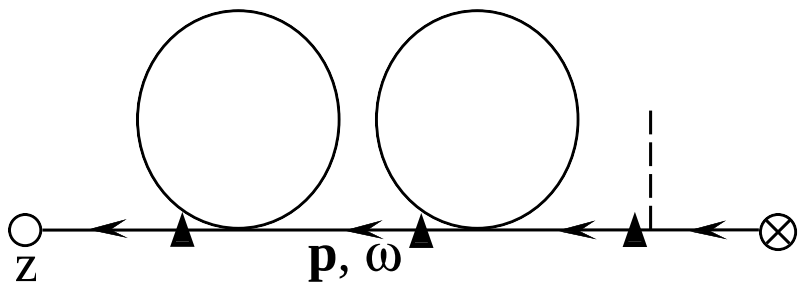

$$= \frac{(s_d u_0)^2}{8} \Big\{ \Big[ \frac{1}{\epsilon^2} + \frac{C_E - 1}{\epsilon} \Big]$$

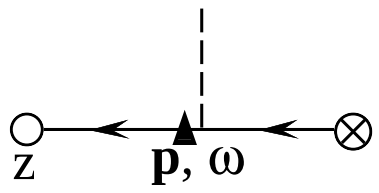

$$- \frac{1}{\epsilon} \frac{F_+ e^{-\kappa_+ z} + F_- e^{-\kappa_- z}}{4\lambda_0(\kappa_+ + \kappa_-)\kappa_+^2 \kappa_-^2} + O(\epsilon^0) \Big\}, \tag{D17}$$

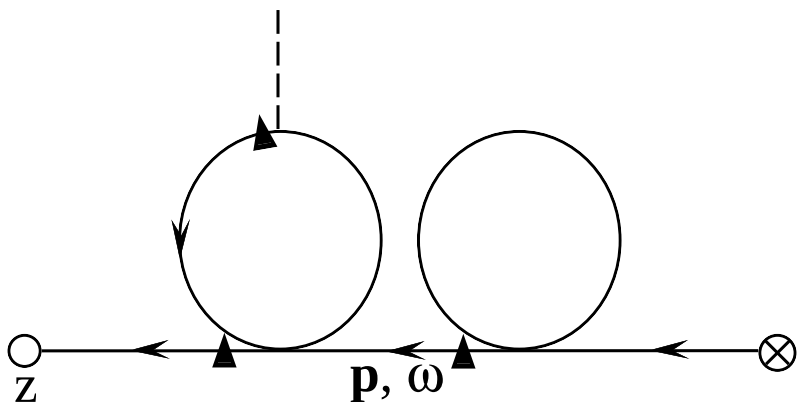

$$= \frac{(s_d u_0)^2}{2}\Bigg\{ -\frac{1}{\epsilon^2} + \frac{1}{\epsilon}\left[ C_E - 2 + \frac{2\kappa_- \ln 2\kappa_+}{\kappa_- - \kappa_+} + \frac{2\kappa_+ \ln 2\kappa_-}{\kappa_+ - \kappa_-} \right]$$

$$+ O(\epsilon^0) \Bigg\}$$ 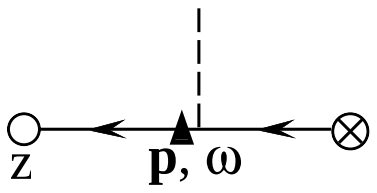

, (D18)

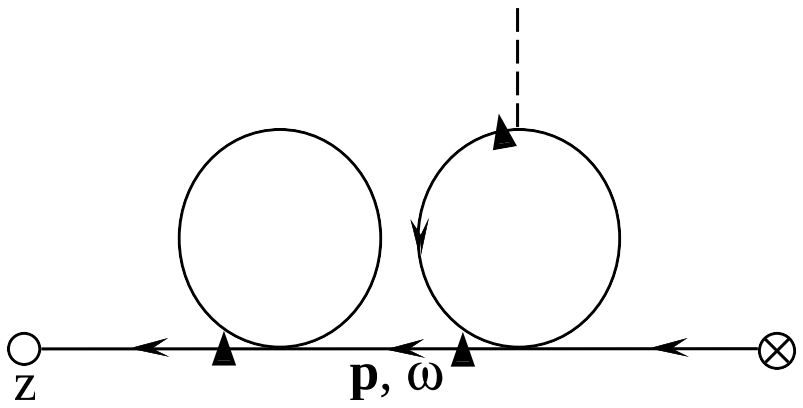

$$= \frac{(s_d u_0)^2}{2}\Bigg\{ \left[ -\frac{1}{\epsilon^2} - \frac{C_E}{\epsilon} \right]$$ 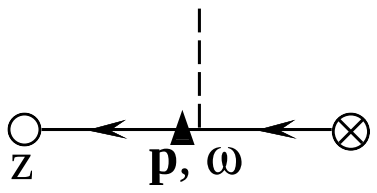

$$+ \frac{1}{\epsilon}\, \frac{F_+ e^{-\kappa_+ z} + F_- e^{-\kappa_- z}}{4\lambda_0(\kappa_+ + \kappa_-)\kappa_+^2 \kappa_-^2} + O(\epsilon^0) \Bigg\}, \qquad \text{(D19)}$$

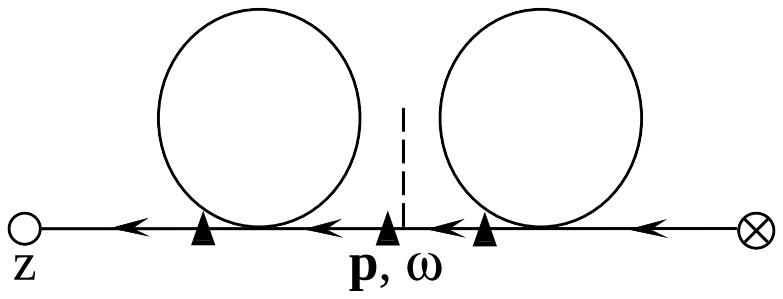

$$= \frac{(s_d u_0)^2}{16}\Bigg\{ \Bigg[ \frac{4}{\epsilon^2} + \frac{1}{\epsilon}\Bigg( 4 - \frac{4\kappa_-}{\kappa_- - \kappa_+} \ln 2\kappa_+$$

$$- \frac{4\kappa_+}{\kappa_+ - \kappa_-} \ln 2\kappa_- \Bigg) \Bigg]$$ 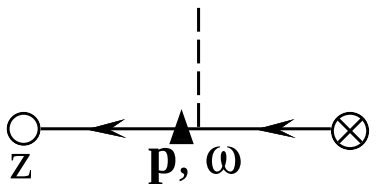

$$- \frac{1}{\epsilon}\, \frac{F_+ e^{-\kappa_+ z} + F_- e^{-\kappa_- z}}{2\lambda_0(\kappa_+ + \kappa_-)\kappa_+^2 \kappa_-^2} + O(\epsilon^0) \Bigg\}, \qquad \text{(D20)}$$

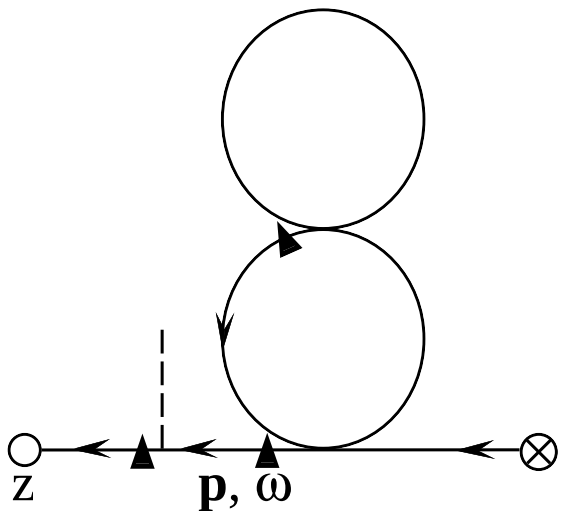

$$= (s_d u_0)^2 \Bigg\{ \left[ -\frac{1}{4\epsilon^2} + \frac{1 - 2C_E}{8\epsilon} \right]$$ 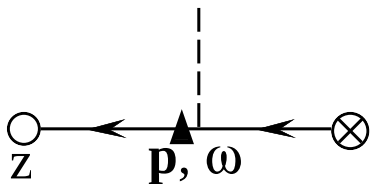

$$+ \frac{1}{4\epsilon}\, \frac{1}{\lambda_0(\kappa_+^2 - \kappa_-^2)\kappa_+^2 \kappa_-^2} \Bigg( \kappa_- e^{-\kappa_+ z} \Big[ (\kappa_+ - \kappa_-)(C_E - 1)$$

$$- \kappa_- \ln 2\kappa_+ + \kappa_+ \ln 2\kappa_- \Big] - \kappa_+ e^{-\kappa_- z} \Big[ (\kappa_- - \kappa_+)(C_E - 1)$$

$$- \kappa_+ \ln 2\kappa_- + \kappa_- \ln 2\kappa_+ \Big] \Bigg) + O(\epsilon^0) \Bigg\}, \qquad \text{(D21)}$$

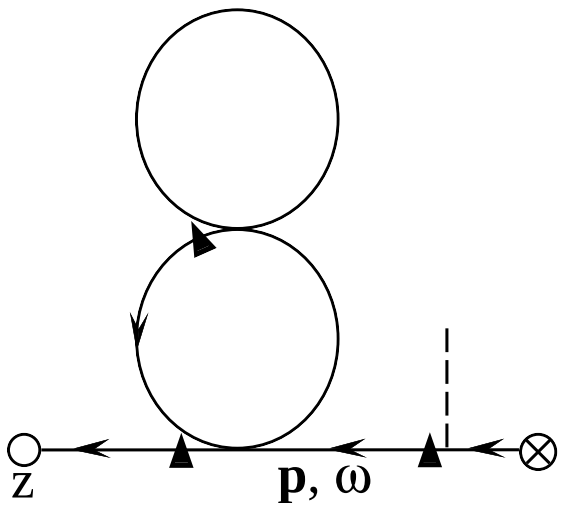

$$= \frac{(s_d u_0)^2}{4}\Bigg\{\left[-\frac{1}{\epsilon^2}+\frac{1-2C_E}{2\epsilon}\right]$$

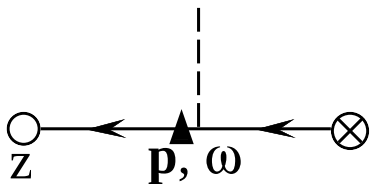

$$+\frac{1}{\epsilon}\,\frac{F_+e^{-\kappa_+ z}+F_-e^{-\kappa_- z}}{4\lambda_0(\kappa_++\kappa_-)\kappa_+^2\kappa_-^2}+O(\epsilon^0)\Bigg\}, \qquad \text{(D22)}$$

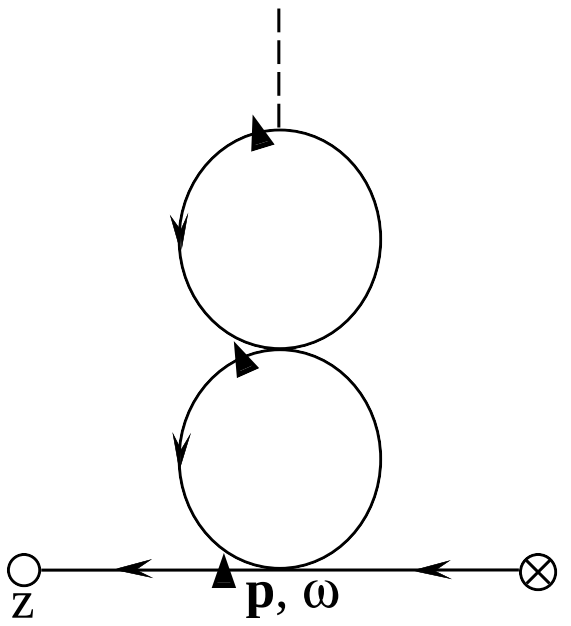

$$= (s_d u_0)^2\Bigg\{\left[\frac{3}{\epsilon^2}+\frac{7+3C_E}{\epsilon}\right]$$

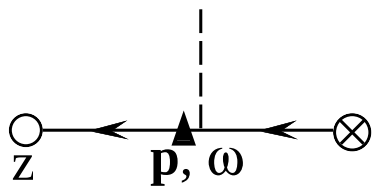

$$+\frac{3}{2\epsilon}\,\frac{G_+e^{-\kappa_+ z}-G_-e^{-\kappa_- z}}{\lambda_0(\kappa_+^2-\kappa_-^2)\kappa_+\kappa_-}+O(\epsilon^0)\Bigg\}, \qquad \text{(D23)}$$

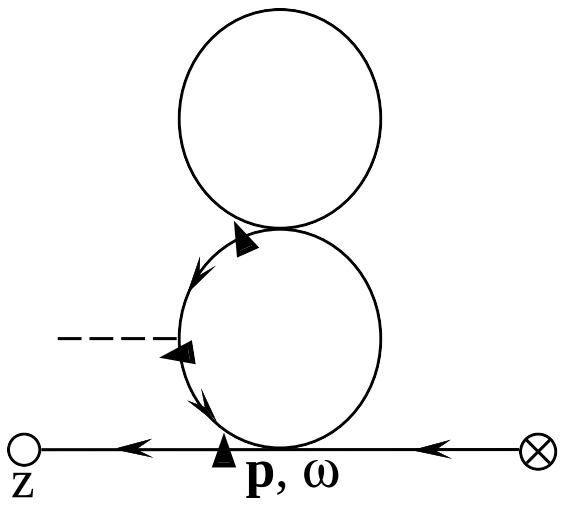

$+$

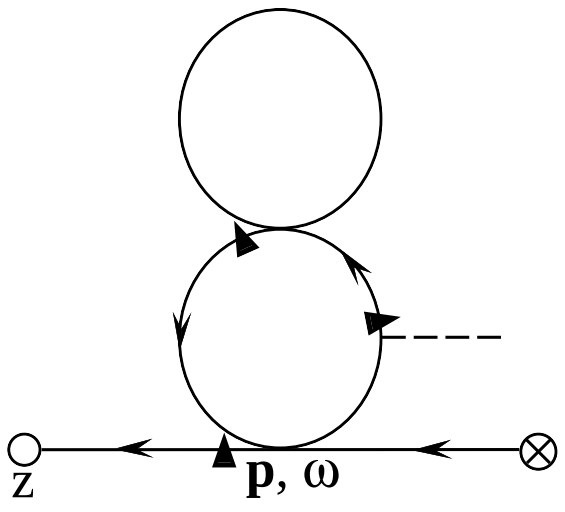

$+$

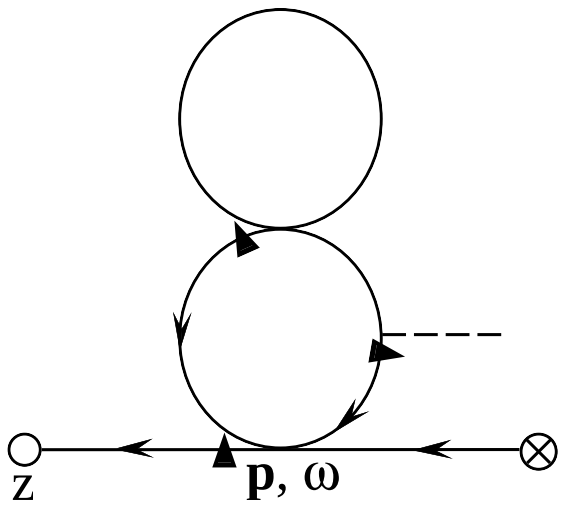

$$= (s_d u_0)^2\Bigg\{\left[-\frac{1}{\epsilon^2}-\frac{3+C_E}{\epsilon}\right]$$

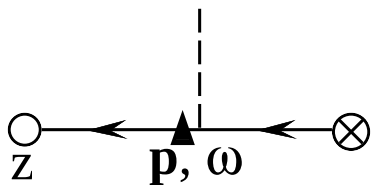

$$-\frac{1}{\epsilon}\,\frac{G_+e^{-\kappa_+ z}-G_-e^{-\kappa_- z}}{2\lambda_0\kappa_+\kappa_-(\kappa_+^2-\kappa_-^2)}+O(\epsilon^0)\Bigg\}, \qquad \text{(D24)}$$

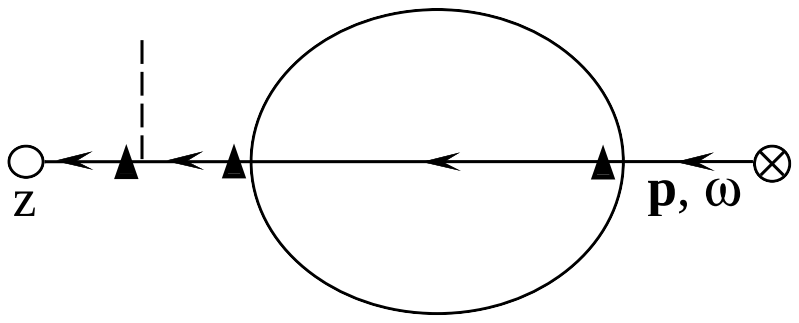

$$= \frac{(s_d u_0)^2}{2}\Bigg\{\left(-\frac{1}{\epsilon^2}+\frac{9+8C_E}{\epsilon}\right)$$

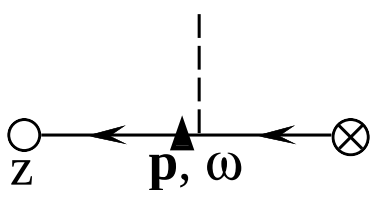

$$+\frac{1}{\epsilon}\Bigg(\frac{1}{96\lambda_0(\kappa_++\kappa_-)\kappa_+^4\kappa_-^4}\Big\{\left[-\kappa_-^4-\kappa_-^3\kappa_++2\kappa_-^2\kappa_+^2-\kappa_-\kappa_+^3-\kappa_+^4\right]\kappa_-e^{-\kappa_+ z}$$

$$+\Big[-\kappa_+^4-\kappa_+^3\kappa_-+2\kappa_+^2\kappa_-^2-\kappa_+\kappa_-^3-\kappa_-^4\Big]\kappa_+e^{-\kappa_-z}\Big\}$$

$$+\frac{1}{\lambda_0(\kappa_+^2-\kappa_-^2)\kappa_+^2\kappa_-^2}\Big\{\Big[(\kappa_+-\kappa_-)(C_E-1)-\kappa_-\ln 2\kappa_++\kappa_+\ln 2\kappa_-\Big]\kappa_-e^{-\kappa_+z}$$

$$-\Big[(\kappa_--\kappa_+)(C_E-1)-\kappa_+\ln 2\kappa_-+\kappa_-\ln 2\kappa_+\Big]\kappa_+e^{-\kappa_-z}\Big\}\Big)+O(\epsilon^0)\Big\}, \qquad \text{(D25)}$$

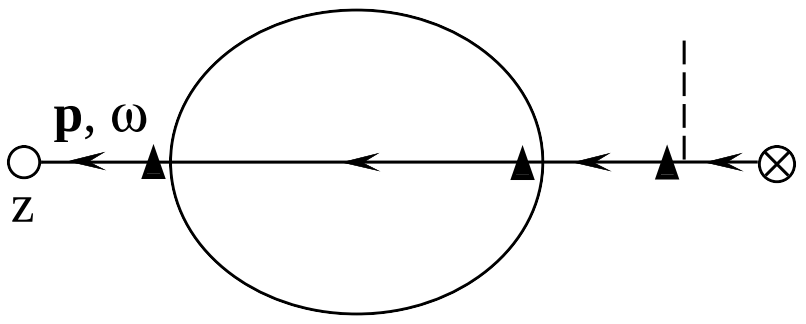

$$=(s_du_0)^2\Big\{\Big(-\frac{1}{2\epsilon^2}-\frac{9+8C_E}{16\epsilon}\Big)$$ 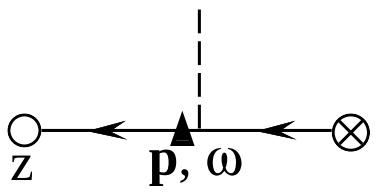

$$+\frac{1}{\epsilon}\,\frac{1}{192\lambda_0(\kappa_++\kappa_-)\kappa_+^4\kappa_-^4}\Big(\Big[z\kappa_+\kappa_-(\kappa_+^2-\kappa_-^2)-\kappa_-(\kappa_+^2+\kappa_-^2)+24\kappa_+^2F_+\Big]\kappa_-^2e^{-\kappa_+z}$$

$$+\Big[z\kappa_-\kappa_+(\kappa_-^2-\kappa_+^2)-\kappa_+(\kappa_+^2+\kappa_-^2)+24\kappa_-^2F_-\Big]\kappa_+^2e^{-\kappa_-z}\Big)+O(\epsilon^0)\Big\}, \qquad \text{(D26)}$$

and

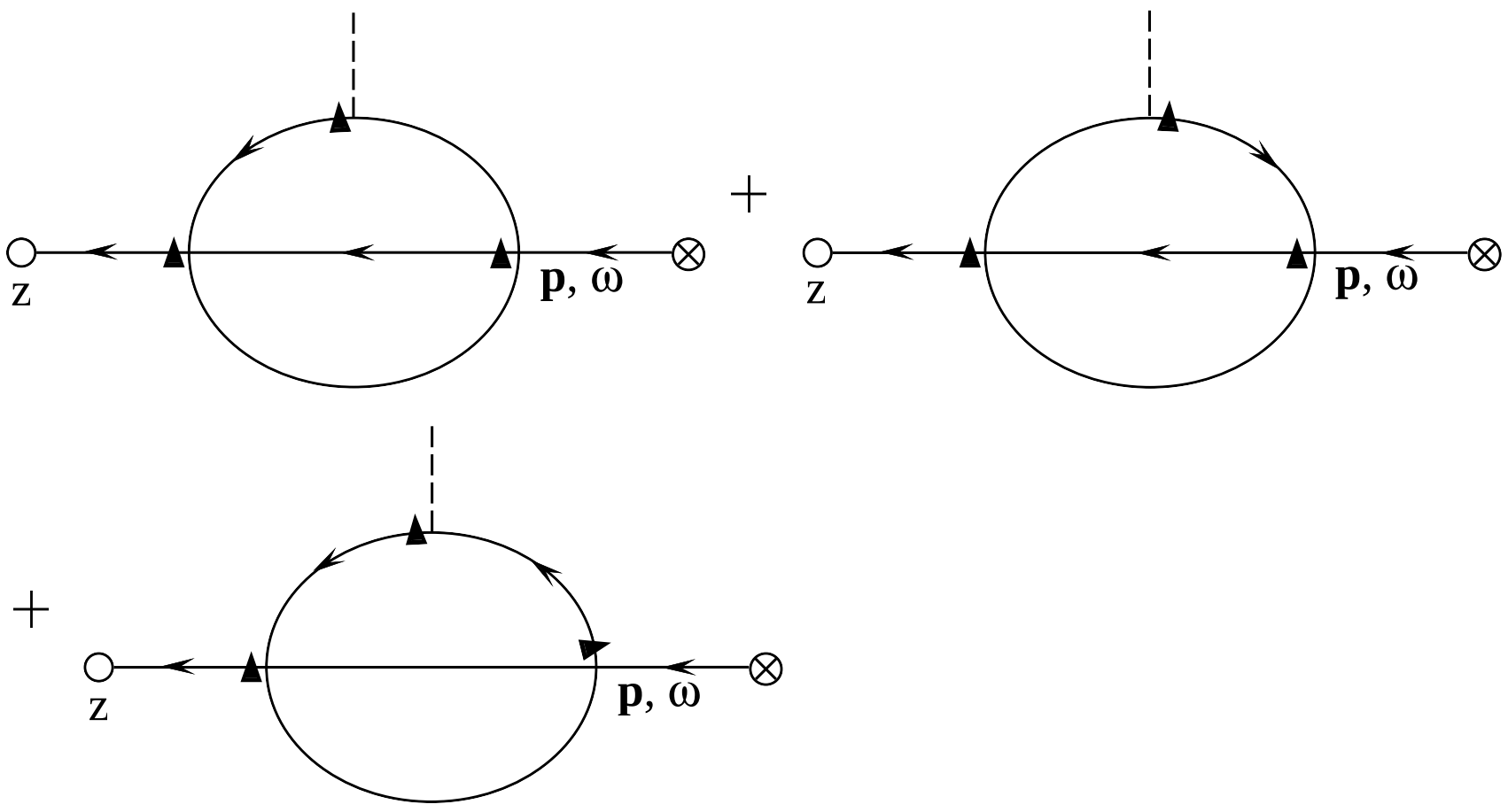

$$=(s_du_0)^2\,\frac{1}{\epsilon}\Big\{\Big(\frac{2}{\epsilon}+3+2C_E+\frac{\pi^2}{3}\Big)$$ 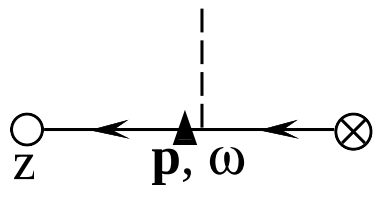

$$+\frac{G_+e^{-\kappa_+z}-G_-e^{-\kappa_-z}}{\lambda_0(\kappa_+^2-\kappa_-^2)\kappa_+\kappa_-}+O(\epsilon)\Big\}. \qquad \text{(D27)}$$

Upon insertion of the relations (C1) for $u_0$ and $\lambda_0$ one obtains the reparametrized graphs. The explicit expressions resulting to order $u^2$ for the zero and one-loop graphs read

$\Big[$ 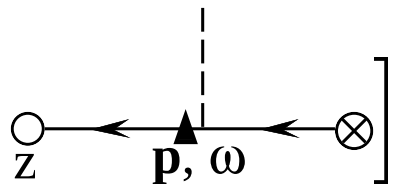 $\Big]_{\mathrm{Rep}}$

$$=\frac{1}{2\lambda(k_++k_-)k_+^2k_-^2}\Big[k_-e^{-k_+z}+k_+e^{-k_-z}\Big]$$

$$+u\Big\{\frac{1}{\epsilon}\frac{1}{2\lambda(k_++k_-)k_+^2k_-^2}\Big[k_-e^{-k_+z}+k_+e^{-k_-z}\Big]+O(\epsilon^0)\Big\}$$

$$+ u^2\Big\{ \frac{1}{\epsilon^2}\frac{1}{\lambda(k_+ + k_-)k_+^2 k_-^2}\Big[k_- e^{-k_+ z} + k_+ e^{-k_- z}\Big]$$
$$+ \frac{1}{\epsilon}\frac{1}{192\lambda k_+^4 k_-^4 (k_+ + k_-)}\bigg(\Big[z k_+ k_-^3 (k_-^2 - k_+^2) + 2k_-^5 + k_-^4 k_+$$
$$+ k_-^3 k_+^2 (11 - 32\pi^2) + k_-^2 k_+^3 + k_- k_+^4\Big] e^{-k_+ z} + \Big[z k_- k_+^3 (k_+^2 - k_-^2)$$
$$+ 2k_+^5 + k_+^4 k_- + k_+^3 k_-^2 (11 - 32\pi^2) + k_+^2 k_-^3 + k_+ k_-^4\Big] e^{-k_- z}\bigg)$$
$$+ O(\epsilon^0)\Big\} + O(u^3)\ , \tag{D28}$$

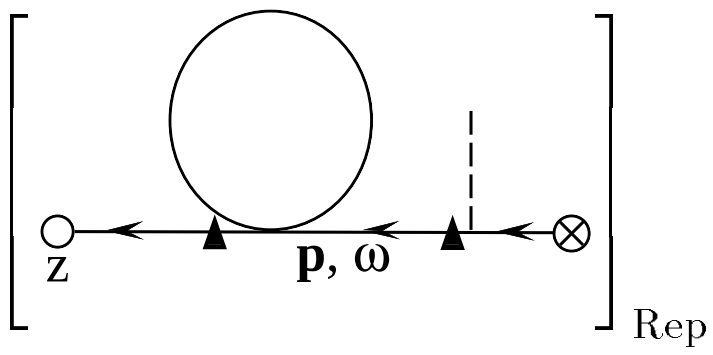

$$= u\Big\{\frac{1}{\epsilon}\frac{1}{4\lambda(k_+ + k_-)k_+^2 k_-^2}\Big[k_- e^{-k_+ z} + k_+ e^{-k_- z}\Big] + O(\epsilon^0)\Big\}$$
$$+ u^2\Big\{\frac{1}{\epsilon^2}\frac{1}{\lambda(k_+ + k_-)k_+^2 k_-^2}\Big[k_- e^{-k_+ z} + k_+ e^{-k_- z}\Big]$$
$$+ \frac{1}{\epsilon}\frac{1}{4\lambda(k_+ + k_-)k_+^2 k_-^2}\Big[(2C_E k_- - \mathcal{F}_+) e^{-k_+ z}$$
$$+ (2C_E k_+ - \mathcal{F}_-) e^{-k_- z}\Big] + O(\epsilon^0)\Big\} + O(u^3)\ , \tag{D29}$$

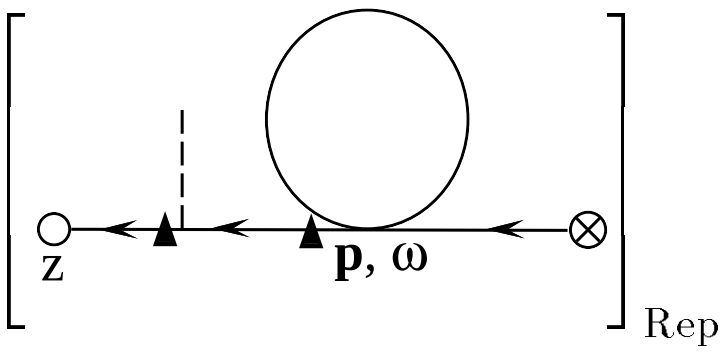

$$= u\Big\{\frac{1}{\epsilon}\frac{1}{4\lambda(k_+ + k_-)k_+^2 k_-^2}\Big[k_- e^{-k_+ z} + k_+ e^{-k_- z}\Big] + O(\epsilon^0)\Big\}$$
$$+ u^2\Big\{\frac{1}{\epsilon^2}\frac{1}{\lambda(k_+ + k_-)k_+^2 k_-^2}\Big[k_- e^{-k_+ z} + k_+ e^{-k_- z}\Big]$$
$$+ \frac{1}{\epsilon}\frac{1}{2\lambda(k_+^2 - k_-^2)k_+^2 k_-^2}\Big[\Big(2k_-^2 \ln 2k_+ - 2k_+ k_- \ln 2k_-$$
$$+ k_-(k_+ - k_-)(2 - C_E)\Big) e^{-k_+ z} - \Big(2k_+^2 \ln 2k_- - 2k_+ k_- \ln 2k_+$$
$$+ k_+(k_- - k_+)(2 - C_E)\Big) e^{-k_- z}\Big] + O(\epsilon^0)\Big\} + O(u^3)\ , \tag{D30}$$

and

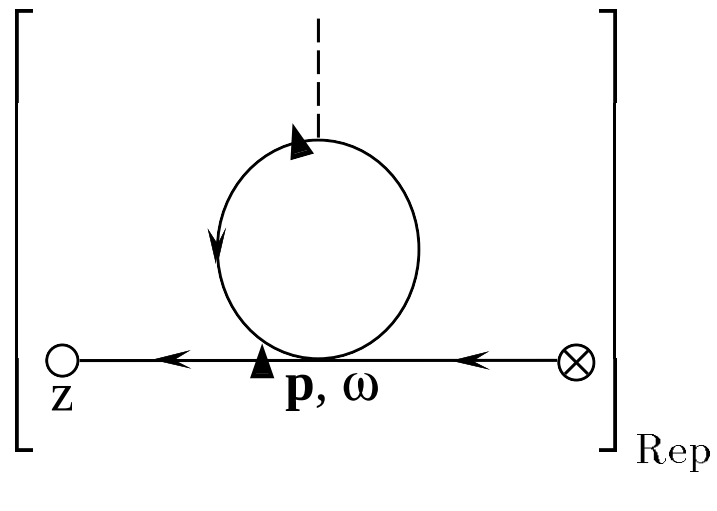

$$= u\Big\{-\frac{1}{\epsilon}\frac{1}{\lambda(k_+ + k_-)k_+^2 k_-^2}\Big[k_- e^{-k_+ z} + k_+ e^{-k_- z}\Big] + O(\epsilon^0)\Big\}$$

$$+ u^2 \Bigg\{ -\frac{1}{\epsilon^2} \frac{4}{\lambda(k_+ + k_-)k_+^2 k_-^2} \Big( k_- e^{-k_+ z} + k_+ e^{-k_- z} \Big)$$
$$-\frac{1}{\epsilon} \frac{2}{\lambda(k_+^2 - k_-^2)k_+^2 k_-^2} \bigg( \Big[ (k_+ - k_-)(2 + C_E) + k_+ \mathcal{G}_+ \Big] k_- e^{-k_+ z}$$
$$- \Big[ (k_- - k_+)(2 + C_E) + k_- \mathcal{G}_- \Big] k_+ e^{-k_- z} \bigg) + O(\epsilon^0) \Bigg\}$$
$$+ O(u^3) \; . \tag{D31}$$

## APPENDIX E: CALCULATION OF SUSCEPTIBILITY SCALING FUNCTIONS

In this appendix we give details of the calculation of the scaling functions $\Psi_A(\mathsf{p}, \mathsf{w})$ and $\Psi_B(\mathsf{p}, \mathsf{w})$ whose $\epsilon$ expansions are given in (45) and (46), respectively. In addition we explain how the asymptotic long-time behavior quoted in (60) and (61) can be deduced. Unlike the other appendices, we now have $\tau > 0$ rather than $\tau = 0$. Accordingly the quantities $\kappa_\pm$ now stand no longer for the $\tau_0 = 0$ expressions (A2) but for

$$\kappa_\pm = \sqrt{p^2 + \frac{\tau_0}{2} \pm \sqrt{\left(\frac{\tau_0}{2}\right)^2 + i\frac{\omega}{\lambda_0}}} \; . \tag{E1}$$

From the zero-loop results of Refs. [16] and [17] one obtains

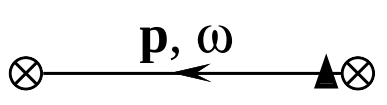
$$\Big|_{\tilde{c}=0} = \frac{(\kappa_+ + \kappa_-)^2 - \tau_0}{2\lambda_0 \kappa_+ \kappa_- (\kappa_+ + \kappa_-)} \tag{E2}$$

and

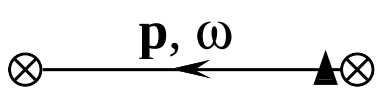
$$\Big|_{\tilde{c}=\infty} = \frac{2(\kappa_+ + \kappa_-)}{\lambda_0[(\kappa_+ + \kappa_-)^2 + \tau_0]} \; . \tag{E3}$$

The one-loop contributions to $\chi_{11}$ involve the vertex graph

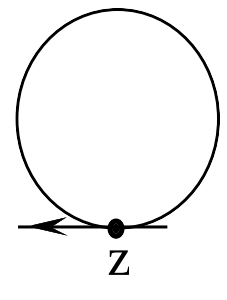
$$= -\frac{\lambda_0 s_d u_0}{2} \tau_0^{1-\frac{\epsilon}{2}} \Big[ \Gamma(\epsilon/2 - 1) + 2\, (z\sqrt{\tau_0})^{-1+\epsilon/2} \, K_{1-\epsilon/2}(2z\sqrt{\tau_0}) \Big] \; , \tag{E4}$$

a result known from the static theory and independent of $\tilde{c}_0$. Here $K_{1-\epsilon/2}(x)$ is a standard Bessel function.

Utilizing this result and performing the necessary integrations, one arrives at

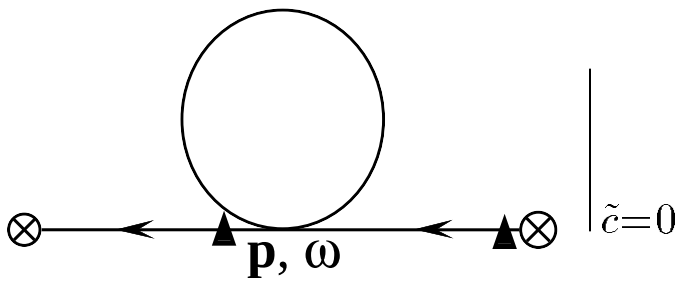

$$
= -\frac{s_d u_0}{8\lambda_0}\,\tau_0^{1-\epsilon/2}\Bigg\{\Gamma\left(\epsilon/2-1\right)\Bigg[2\,\frac{1-\frac{\tau_0^2}{\left(\kappa_+^2-\kappa_-^2\right)^2}}{\kappa_+\kappa_-(\kappa_+ + \kappa_-)} + \frac{1}{2\kappa_+^3}\left(1+\frac{\tau_0}{\kappa_+^2-\kappa_-^2}\right)^2
$$

$$
+ \frac{1}{2\kappa_-^3}\left(1+\frac{\tau_0}{\kappa_-^2-\kappa_+^2}\right)^2\Bigg] + \frac{\sqrt{\pi}\,2^{2-\epsilon}\Gamma(\epsilon-1)}{\Gamma[(1+\epsilon)/2]}\Bigg[\frac{\left(1+\frac{\tau_0}{\kappa_+^2-\kappa_-^2}\right)^2}{\kappa_+^2(\kappa_+ + \sqrt{\tau_0})}\,F_\epsilon\!\left(\frac{\kappa_+ - \sqrt{\tau_0}}{\kappa_+ + \sqrt{\tau_0}}\right)
$$

$$
+ \frac{\left(1+\frac{\tau_0}{\kappa_-^2-\kappa_+^2}\right)^2}{\kappa_-^2(\kappa_- + \sqrt{\tau_0})}\,F_\epsilon\!\left(\frac{\kappa_- - \sqrt{\tau_0}}{\kappa_- + \sqrt{\tau_0}}\right) + 4\,\frac{1-\frac{\tau_0^2}{\left(\kappa_+^2-\kappa_-^2\right)^2}}{\kappa_+\kappa_-(\kappa_+ + \kappa_- + 2\sqrt{\tau_0})}\,F_\epsilon\!\left(\frac{\kappa_+ + \kappa_- - 2\sqrt{\tau_0}}{\kappa_+ + \kappa_- + 2\sqrt{\tau_0}}\right)\Bigg]\Bigg\}
\tag{E5}
$$

and

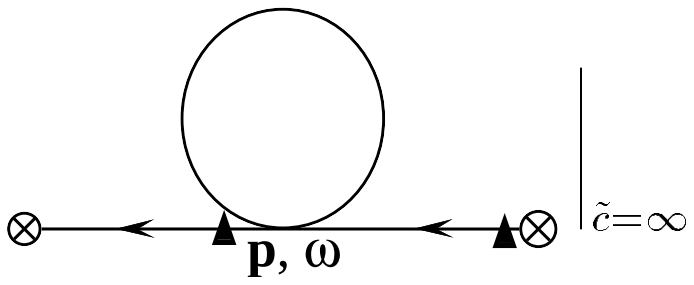

$$
= -\frac{s_d u_0}{2\lambda_0}\Bigg\{\Gamma(\epsilon/2-1)\,\tau_0^{1-\epsilon/2}\,\frac{(\kappa_+ + \kappa_-)^2(\kappa_+^2 + 6\kappa_+\kappa_- + \kappa_-^2) - 2\tau_0(\kappa_+ + \kappa_-)^2 + \tau_0^2}{2\kappa_+\kappa_-(\kappa_+ + \kappa_-)\left[(\kappa_+ + \kappa_-)^2 + \tau_0\right]^2}
$$

$$
+ \frac{\sqrt{\pi}\,2^{2-\epsilon}\tau_0^{1-\epsilon/2}\,\Gamma(\epsilon-1)}{\Gamma[(1+\epsilon)/2]\,(\kappa_+ - \kappa_-)^2\left[(\kappa_+ + \kappa_-)^2 + \tau_0\right]^2}\Bigg[\frac{(\kappa_+^2 - \kappa_-^2 + \tau_0)^2}{\kappa_+ + \sqrt{\tau_0}}\,F_\epsilon\!\left(\frac{\kappa_+ - \sqrt{\tau_0}}{\kappa_+ + \sqrt{\tau_0}}\right)
$$

$$
+ \frac{(\kappa_-^2 - \kappa_+^2 + \tau_0)^2}{\kappa_- + \sqrt{\tau_0}}\,F_\epsilon\!\left(\frac{\kappa_- - \sqrt{\tau_0}}{\kappa_- + \sqrt{\tau_0}}\right)
$$

$$
- \frac{4(\kappa_+^2 - \kappa_-^2 + \tau_0)(\kappa_-^2 - \kappa_+^2 + \tau_0)}{\kappa_+ + \kappa_- + 2\sqrt{\tau_0}}\,F_\epsilon\!\left(\frac{\kappa_+ + \kappa_- - 2\sqrt{\tau_0}}{\kappa_+ + \kappa_- + 2\sqrt{\tau_0}}\right)\Bigg]\Bigg\},
\tag{E6}
$$

where $F_\epsilon$ is the hypergeometric function

$$
F_\epsilon(x) \equiv {}_2F_1\!\left(1, \frac{3-\epsilon}{2}; \frac{1+\epsilon}{2}; x\right).
\tag{E7}
$$

We use the expansion

$$
F_\epsilon(x) = \frac{1+x}{(1-x)^2} - \epsilon\,\frac{2\sqrt{x}}{(1-x)^2}\,\ln\frac{1+\sqrt{x}}{1-\sqrt{x}} + O(\epsilon^2)\;.
\tag{E8}
$$

With the aid of the the reparametrizations (13a)–(15b) and the perturbative results for the $Z$-factors quoted in (31), the renormalized susceptibilities $\chi_{11,R}(\tilde{c}=0)$ and $\chi_{11,R}(\tilde{c}=\infty)$ then follow in a straightforward fashion. Evaluating these at $u = u^* = \epsilon/3 + O(\epsilon)$, the value of the infrared-stable fixed point, leads to the results for $\Psi_B$ and $\Psi_A$ presented in (46) and (45).

## APPENDIX F: LONG-TIME BEHAVIOR OF SUSCEPTIBILITY SCALING FUNCTIONS

We wish to determine the asymptotic behavior of the zero-loop scaling functions

$$\Psi^0_{A,B}(\mathsf{p},\mathsf{t}) = \frac{1}{2\pi}\int_{-\infty}^{\infty} d\mathsf{w}\,\hat{\Psi}^0_{A,B}(\mathsf{p},\mathsf{w})\,e^{-i\mathsf{w}\mathsf{t}} \tag{F1}$$

as $\mathsf{t}\to\infty$. The functions $\Psi^0_{A,B}(\mathsf{p},\mathsf{w})$ have a branch cut for $\mathrm{Im}\,\mathsf{w} < -\mathsf{p}^2(1-\mathsf{p}^2)$ and no pole. We deform the integration path as shown in Fig. 4. The contributions from $I_1$, $I_3$, and $I_5$ vanish. The contributions from $I_2$ and $I_4$ add up to

$$\Psi^0_{A,B}(\mathsf{p},\mathsf{t}) = -\frac{1}{\pi}\int_{\mathsf{p}^2(1-\mathsf{p}^2)}^{\infty} dx\,\mathrm{Im}\hat{\Psi}^0_{A,B}(\mathsf{p},-ix)\,e^{-x\mathsf{t}}. \tag{F2}$$

Expansion of $\mathrm{Im}\Psi^0_{A,B}$ in $y \equiv x - \mathsf{p}^2(1-\mathsf{p}^2)$ yields

$$\mathrm{Im}\hat{\Psi}^0_{A}[\mathsf{p},-i(y+\mathsf{p}^2+\mathsf{p}^4)] = -\frac{\sqrt{1+2\mathsf{p}^2}}{(1+\mathsf{p}^2)^2\,y^{1/2}}\,e^{-(\mathsf{p}^2+\mathsf{p}^4)\mathsf{t}}\left[\mathsf{p}^2 - \frac{(5\mathsf{p}^2+5\mathsf{p}^6-2)\,y}{2\,(1+2\mathsf{p}^2)^2\,(1+\mathsf{p}^2)^2} + O(y^2)\right] \tag{F3}$$

and

$$\mathrm{Im}\hat{\Psi}^0_{A}[\mathsf{p},-i(y+\mathsf{p}^2+\mathsf{p}^4)] = -\sqrt{1+2\mathsf{p}^2}\,\frac{e^{-(\mathsf{p}^2+\mathsf{p}^4)\mathsf{t}}}{y^{1/2}}\left[\mathsf{p}^2 + \frac{(2+\mathsf{p}^2)\,y}{2\,(1+2\mathsf{p}^2)^2} + O(y^2)\right]. \tag{F4}$$

Upon substitution of these expansions into (F2), the integrations can be performed. This gives the asymptotic forms (60) and (61).

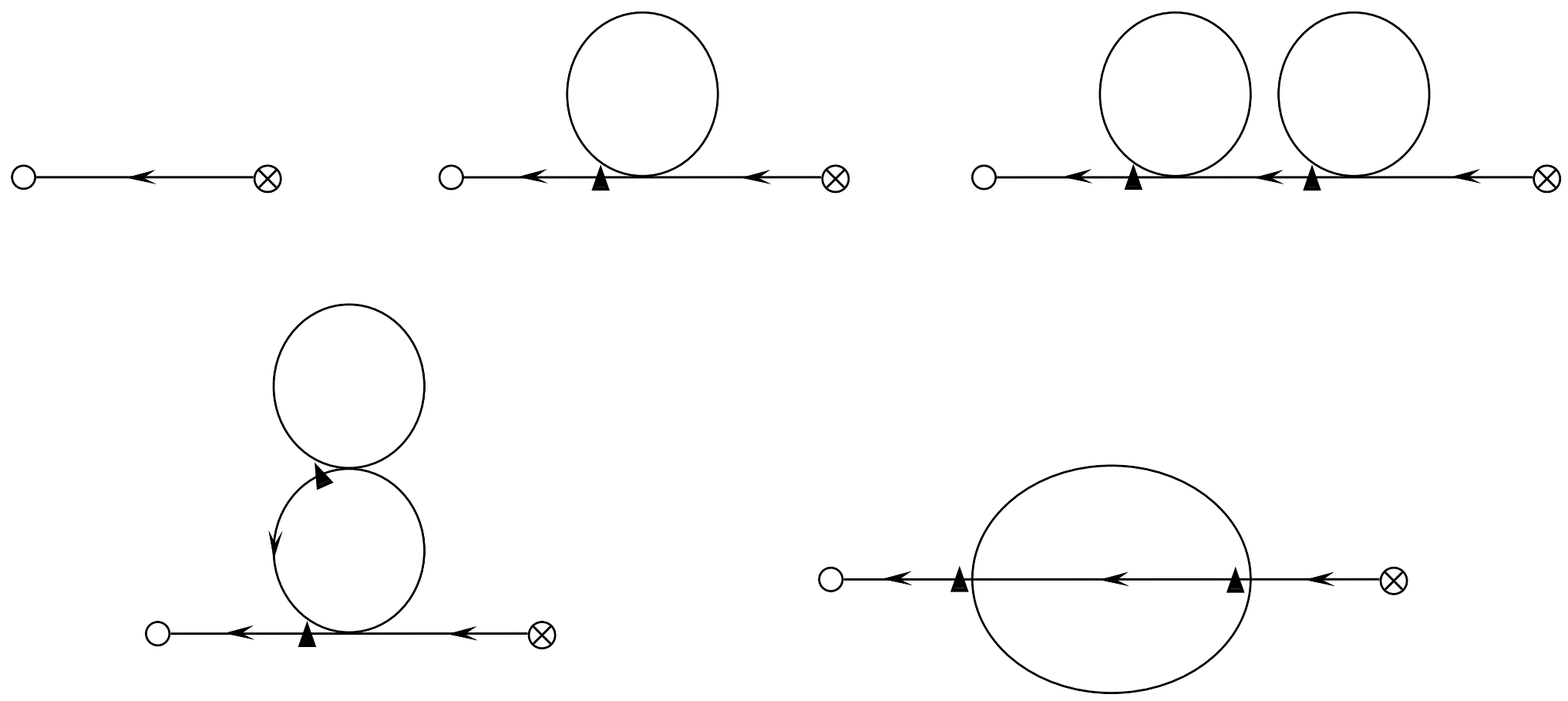

FIG. 1. Graphs of $G^{(0,1;1,0)}$.

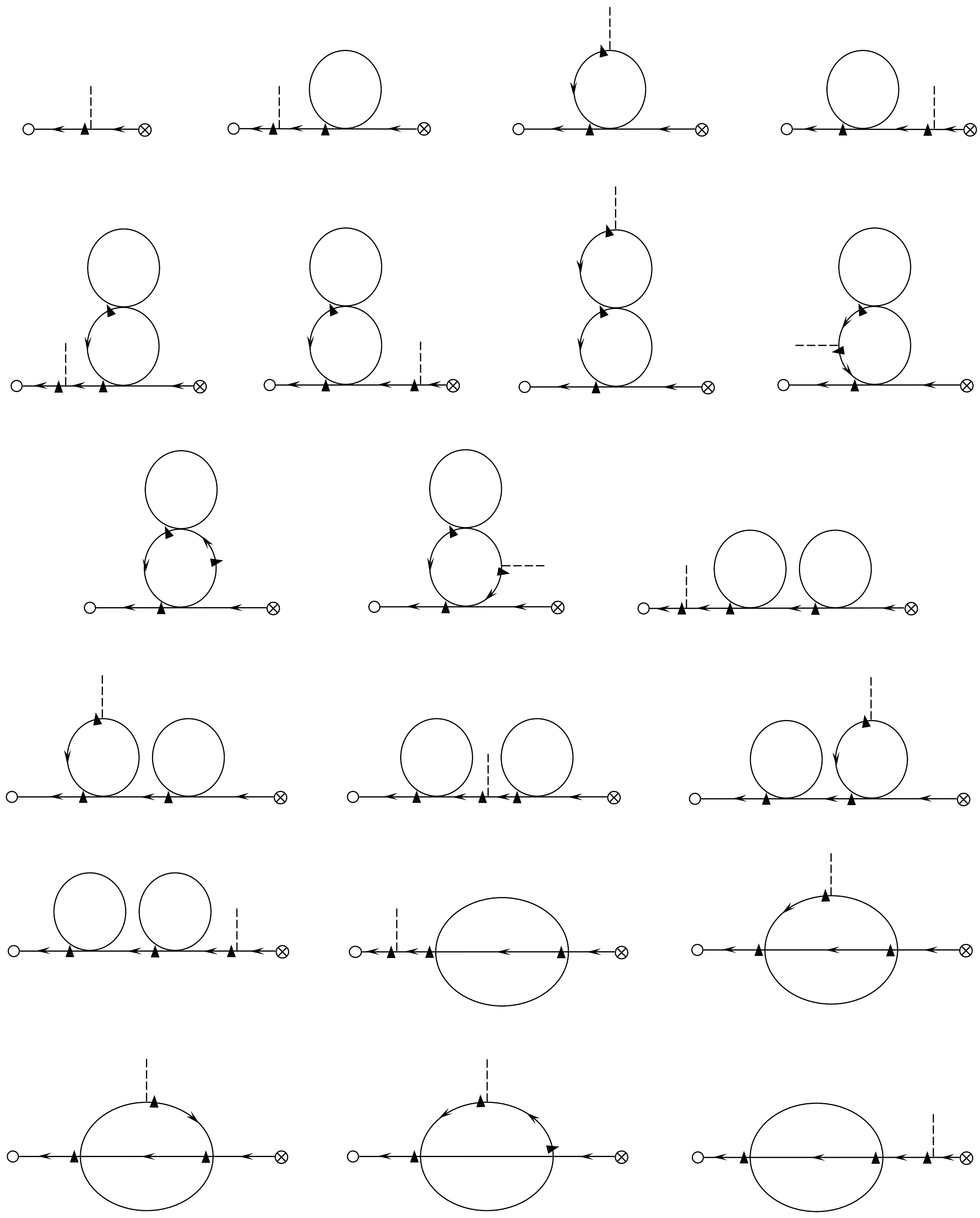

FIG. 2. Graphs of $G^{(0,1;1,0;1)}$.

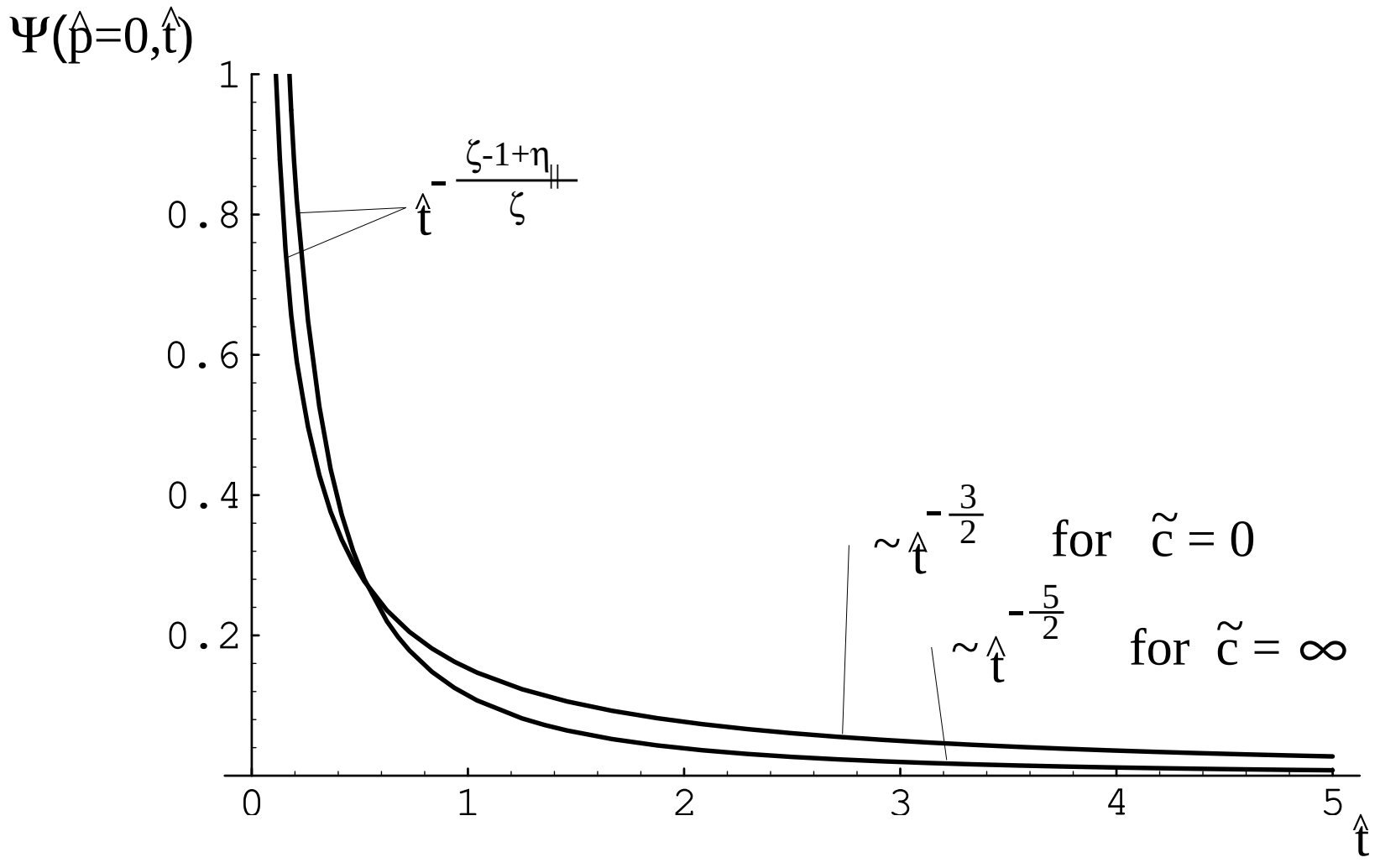

FIG. 3. Scaling functions $\Psi_A$ and $\Psi_B$ of $\chi_{11}$, extrapolated to $d = 3$.

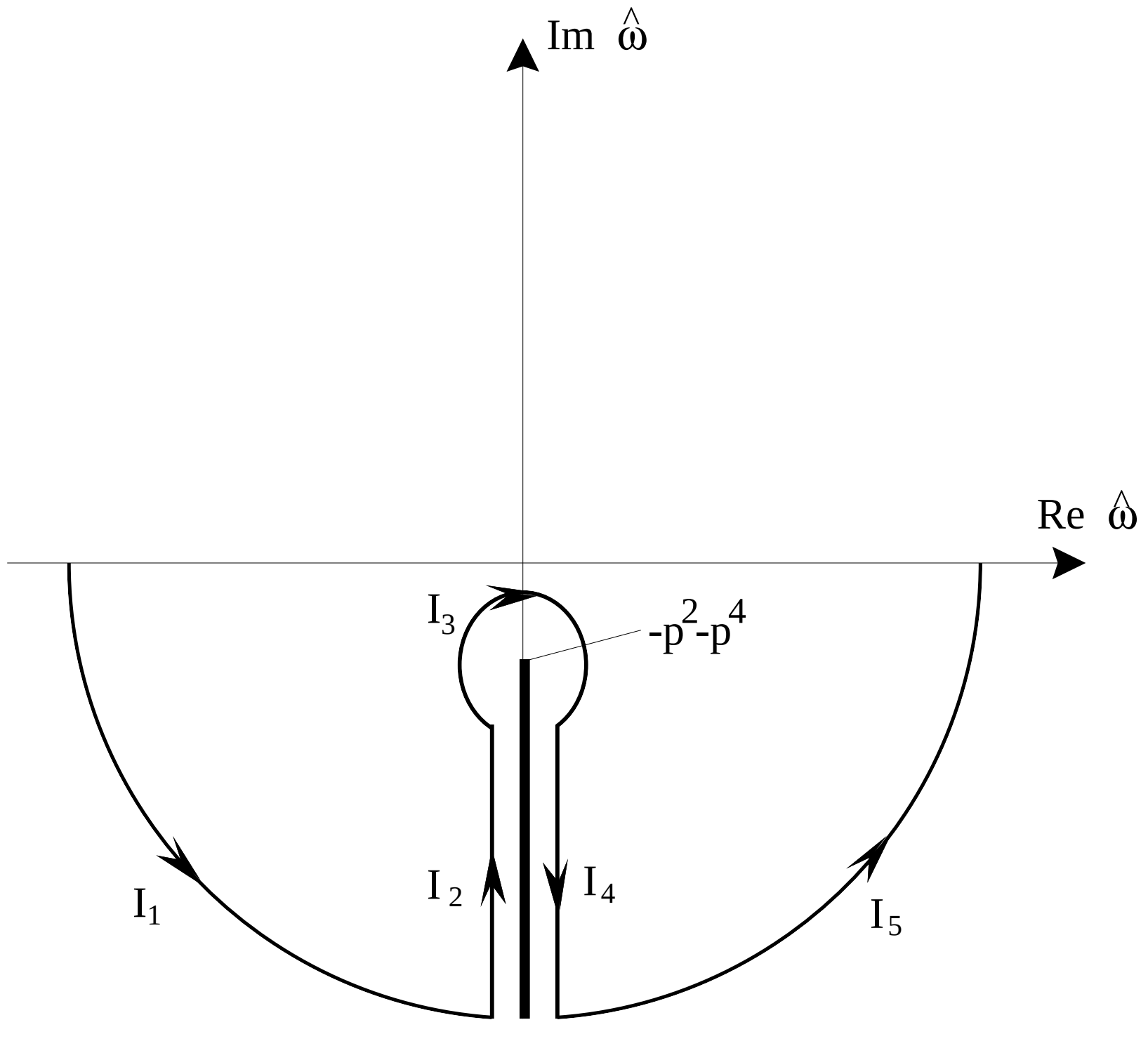

FIG. 4. Integration path used to calculate $\Psi_A$ and $\Psi_B$.